\documentclass[preprint2]{proto}
\usepackage{amsmath}
\usepackage{times}
\usepackage{xcolor}
\usepackage{graphicx}
\usepackage{wasysym}
\usepackage{float}
\usepackage{dblfloatfix}
\usepackage[T1]{fontenc}

\graphicspath{{./}{Figures/}}

\def\msun{{\rm\,M_\odot}}

\newcommand{\pfrac}[2]{ \left( \displaystyle\frac{#1}{#2} \right) }

\begin{document}

\title{\textbf{\LARGE The physical and chemical processes in protoplanetary disks: constraints on the composition of comets}}

\author {\textbf{\large Yuri Aikawa}}
\affil{\small\em Department of Astronomy, The University of Tokyo}

\author {\textbf{\large Satoshi Okuzumi}}
\affil{\small\em Department of Earth and Planetary Sciences, Institute of Science Tokyo}

\author {\textbf{\large Klaus Pontoppidan}}
\affil{\small\em Jet Propulsion Laboratory}

\begin{abstract}

\begin{list}{ } {\rightmargin 1in}
\baselineskip = 11pt
\parindent=1pc
{\small 

We review the recent observations of protoplanetary disks together with relevant theoretical studies with an emphasis on the evolution of volatiles.
In the last several years {\it Atacama Large Millimeter/submillimeter Array} (ALMA) provided evidence of grain growth, gas-dust decoupling, and sub-structures such as rings and gaps in the dust continuum. Molecular line observations revealed radial and vertical distributions of molecular abundances and also provided significant constraints on the gas dynamics such as turbulence. While sub-millimeter and millimeter observations mainly probe the gas and dust outside the radius of several au, ice and inner warm gas are investigated at shorter wavelengths.
Gas and dust dynamics are key to connect these observational findings. One of the emerging trends is in-homogeneous distributions of elemental abundances, most probably due to dust-gas decoupling.

}
\end{list}
\end{abstract}


\section{INTRODUCTION}
\label{sec:intro}

Planetary systems, including our Solar system, are formed in protoplanetary disks, which are circumstellar disks around pre-main-sequence stars. Since COMET II, new astronomical instruments, such as ALMA and {\it Herschel Space Observatory}, made dramatic progress in the observation of protoplanetary disks. In particular, ALMA made it possible to observe the thermal emission of dust grains with a spatial resolution of 5 au in the disks in the solar neighborhood (e.g. the Ophiuchus star-forming region). The discovery of the ring-gap structure stimulated theoretical studies on grain growth and planetesimal formation processes. Various gas molecules such as CO, on the other hand, are observed with a spatial resolution of $\sim 15$ au. Observations of these molecules provide an important clue to unveil the composition of ice, which, together with dust, is the ingredient of comets. In this chapter, we review the progress of these observational and theoretical studies of protoplanetary disks, which serve as references to investigate the comet formation in the Solar nebula.

\begin{figure*}[t]
\centering
\resizebox{0.8\hsize}{!}{\includegraphics{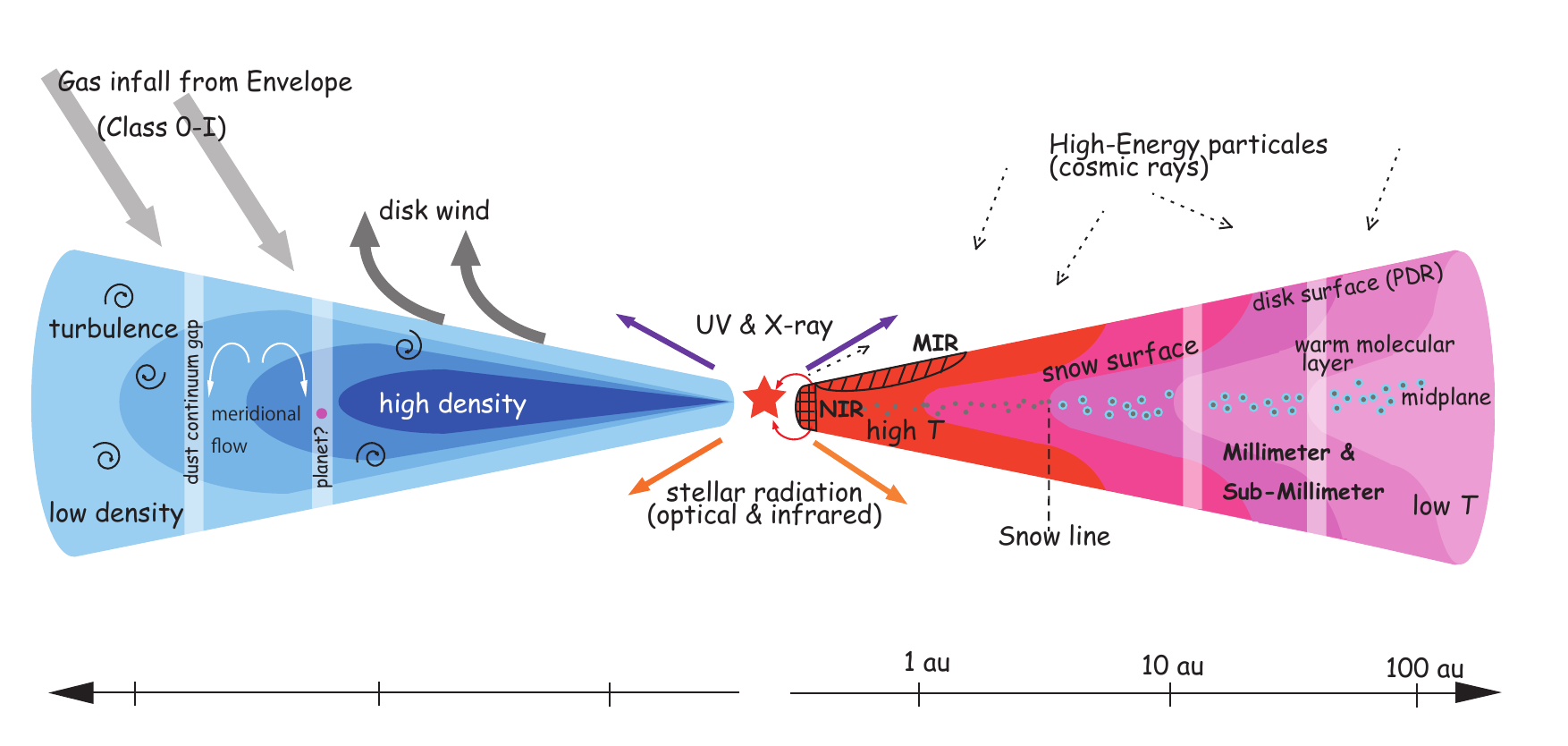}}
\caption{Schematic view of the disk structure. The physical structures and processes are shown on the left: density distribution (blue contour), weak turbulence, disk wind, gaps seen in dust continuum emission, and meridional flow. In early disk-forming stages (Class 0 and I), we also have gas accretion from the envelope. The temperature distribution (red contour) is depicted on the right, together with the basic chemical structures: disk surface irradiated by UV radiation, warm molecular layer, and midplane. Snow surface is defined as the boundary outside which the dust temperature is lower than the sublimation temperature of a volatile molecule. Snow line is the intersection between the snow surface and the disk midplane. Infrared observations trace hot regions at the inner radius and disk surface, while millimeter and sub-millimeter observations probe the outer ($\ge 5$ au) regions.}
\label{fig:schematic}
\end{figure*}

Before proceeding to the main sections, we briefly explain the process of star formation in order to set up the stage and to clarify the relationship with the chapter by Bergin et al. Low-mass stars like the Sun are formed by the gravitational collapse of a molecular cloud core. Since the core has non-zero angular momentum as a whole, the central star and its circumstellar disk are formed simultaneously. In the early stages, the star and disk are deeply embedded in dense gas, which we call envelope. Since the envelope is cold, the spectral energy distribution (SED) of the object has a peak in the far infrared. The objects in this evolutionary stage are called Class 0. In spite of its low temperature at the outermost radius of the envelope, the temperature of the forming disk would be high due to the high mass accretion rate inside the disk (i.e. release of gravitational energy), and shock heating caused by accretion from the envelope to the disk \citep[e.g.][see also Fig. \ref{fig:schematic} and \S \ref{sec:temperature}]{Offner2011}.
While the envelope gas accretes to the central star via disk, the high-velocity jet and outflow are launched from the vicinity of the central star. Eventually, the envelope gas decreases as it falls to the disk and star, or it is swept by the jet and outflow. When the central star (i.e. in near-infrared wavelengths) becomes apparent in the SED, which is still dominated by mid- to far-infrared though, the objects are called Class I. Eventually, when the envelope gas is dissipated, the central star becomes a T Tauri star and the disk becomes a protoplanetary disk, which is also called Class II disk.
More detailed explanations and quantitative definitions of Class 0, I, and II are found in \citet{evans2009}.

While it is not easy to determine the age of pre-main-sequence stars, the typical age of the Class II objects is a few $10^6$ years. Statistical observations (i.e. number counts) indicate that the lifetime for Class 0 and Class I is $\sim 0.1-0.2$ Myr and $\sim 0.4-0.5$ Myr, respectively \citep[e.g.][]{white2007,evans2009}.
In order to set Solar system formation in an astronomical context, these timescales are compared with the range of absolute ages (i.e. formation interval) of Calcium and Aluminium Inclusions (CAIs) and chondrules in meteorites. 
CAIs have the highest condensation temperature among minerals in chondrites and are thus considered to be formed in the hottest stage and/or region.
The formation interval of CAIs is estimated to be 0.16 Myr by the isotope dating, which is comparable to the lifetime of Class 0 stage. The age of the chondrules, on the other hand, varies over a few Myr, which coincides with the lifetime of Class II disks \citep{Connelly2012}.  While Class II disks have been considered to be the birthplace of planetary systems, a ring-gap structure of dust thermal emission is recently found in some Class I disks as well. It suggests that planetary system formation may start early (see \S 4). In summary, the Solar nebula may correspond to the whole evolutionary stages from Class 0, I, and to II.

Having said that, in this chapter, we focus our attention on Class II disks. Chondrule fragments found in the sample return mission of Comet Wild 2 indicate the formation of comets during or after the chondrule formation \citep{Bridges2012}, i.e. in Class II disks. The physical and chemical processes and structures in Class II disks are better understood as depicted in Fig. \ref{fig:schematic} compared with the younger disks, which need more careful analysis to be distinguished from the accreting envelope gas and outflows \citep[e.g.][]{WISH2021}. 
In \S 2, we describe the physical structure and processes in disks such as temperature distributions and gas dynamics, which are of fundamental importance for chemistry and comet formation in disks. Since comets contain significant amounts of volatiles such as water ice, they are considered to be formed in low-temperature regions outside the snowline. In \S 3 we review the basic chemical structure of the disk and recent observations of molecular gas and ice at various wavelengths, as well as the studies on isotope fractionation in disks. The chemical properties of disk material provide a reference for investigations of the formation and evolution of comets based on their volatile composition and isotope ratios.
In \S 4 we overview two outstanding issues on disk chemistry; the chemical evolution of volatiles from earlier phases (Class 0 and I) to Class II disks, and how and what we can learn about volatiles in solids from line observations of disks and central stars.

\section{\textbf{Physical processes in protoplanetary disks}}
\label{sec:nextsection}


In this section, we describe the basic physical structure and relevant observational results. The radial temperature distribution in disks determines snowlines, outside of which specific volatile species will be in ice to be incorporated into comets. Vertical temperature structure is also important to interpret the disk observations. The major heating source is stellar irradiation and gravitational energy released by the mass accretion towards the central star. Both the stellar luminosity and mass accretion rate are expected to vary with time, which results in temporal variation of the snowlines. Mass accretion, in turn, is determined by the angular momentum transfer within the disk, which could be caused by turbulence and/or disk winds (Fig.\ref{fig:schematic}). We note that angular momentum transfer and turbulence will also determine how the ingredients of comets, ices and grains, are mixed and distributed within the disk. We thus start this section with gas dynamics. Later in this section, we also provide the current overview of dust observations. Observations indicate that dust grains have grown at least to 100 $\mu$m, and possibly to larger sizes. As grains grow, they decouple from the gas. Large grains, called pebbles, lose or gain angular momentum via gas drag to radially migrate, and are concentrated at a local gas pressure maximum. Such dust rings (and gaps) are found in many disks in the last several years.

\subsection{Gas dynamics}\label{sec:gas}

Pre-main-sequence stars accrete material from their surrounding disks. For solar-mass pre-main-sequence stars, the mass accretion rates estimated from observations are on average $\sim 10^{-8}~\rm \msun~\rm yr^{-1}$ at a stellar age $t$ of $\sim 1~\rm Myr$ and crudely scale inversely with $t$ \citep{Hartmann2016}. Some pre-main-sequence stars show months-long and decades-long luminosity eruptions, called EX Ori and FU Ori outbursts, with estimated accretion rates of $\sim 10^{-7} M_\odot~\rm yr^{-1}$ and $\sim 10^{-5}$--$10^{-4} M_\odot~\rm yr^{-1}$, respectively \citep{Hartmann2016,Fischer23}.

The observed stellar accretion indicates that the inner region of disks loses angular momentum.
What mechanisms are responsible for the angular momentum transport is a long-standing question in the study of protoplanetary disk evolution.
Classically, the accretion of protoplanetary disks as well as other astrophysical accretion disks was attributed to outward angular momentum transport within the disks by turbulence \citep{Lynden-Bell1974}. However, as we detail below, neither theory nor observations support the picture that protoplanetary disks are strongly turbulent everywhere. 

On the theoretical side, it was previously thought that the magnetorotational instability (MRI;  \citealt{Balbus1991}) is a major cause of protoplanetary disk turbulence.
The MRI is most likely to operate in the innermost disk regions where the gas is hot ($\ge 1000$ K) and thermally well-ionized \citep{Gammie1996,Desch2015}. However, farther out in the disks, the gas is only poorly ionized and the magnetohydrodynamics (MHD) is strongly subject to magnetic diffusion \citep{Sano2000,Ilgner2006,Wardle2007,Bai2011}. Recent theoretical studies  \citep{Bai2011,BaiStone2013,Simon2013a, Simon2013b,Lesur2014,Gressel2015,Bai2017} have shown that while Ohmic diffusion suppresses the MRI near the disk midplane, ambipolar diffusion (which is a type of magnetic diffusion occurring in low-density regions) also suppresses the MRI well above the midplane. In the inner few au of the disks, the combined effect of the Ohmic and ambipolar diffusion can quench the MRI at all heights \citep{Bai2013}.

From the observational side, direct constraints on disk turbulence strength have been obtained from measurements of nonthermal Doppler broadening of molecular emission lines \citep{Hughes2011,Guilloteau2012}. Recently, there have been several attempts to detect disks' nonthermal gas motions using ALMA, but most of them have resulted in non-detection with upper limits on the nonthermal velocity dispersion of several to ten \% of the local sound speed \citep{Flaherty2015,Flaherty2017,Flaherty2018,Teague2016,TeagueHenning2018}. To date, the disks around DM Tau \citep{Flaherty2020} and IM Lup \citep{Paneque-Carreno24, Flaherty24} are the only cases for which turbulent motion has been detected with ALMA , albeit at a few gas scale heights above the midplane.
There are indirect constraints on turbulence strength at the midplane from the morphology or azimuthal emission variation of dust rings and gaps seen in ALMA millimeter continuum images (\citealt{Pinte2016,Dullemond2018,Rosotti2020,Doi2021}; see Sect.~\ref{sec:substructures} for more detail about disk substructures found by millimeter observations).
Some rings show an indication of strong vertical settling or radial dust concentration, suggesting that the turbulence at the midplane of the disks is too weak to diffuse the dust particles. However, constraining the level of turbulence quantitatively from these observations requires additional constraints on the dust particle size because large dust particles may settle and concentrate even in the presence of strong turbulence.

While the turbulence is found to be weak, we still need angular momentum transport to account for the mass accretion from the disk to the central star. Various mechanisms are proposed and investigated.
The Keplerian shear of the gas disk amplifies large-scale horizontal magnetic fields even when Ohmic diffusion suppresses the MRI \citep{Turner2008}. The Keplerian shear, when coupled to the Hall drift of magnetic fields, also leads to exponential amplification of horizontal magnetic fields, the phenomenon called the Hall-shear instability \citep{Kunz2008}, which becomes prominent in inner disk regions \citep{Bai2014,Bai2017,Lesur2014}. Importantly, MHD simulations show that these mechanisms produce {\it coherent} horizontal magnetic fields with no appreciable level of turbulence. In other words, they induce angular momentum transport leading to large-scale disk accretion but would not cause small-scale mixing of disk material. 

Large-scale magnetic fields threading the disks induce another important dynamical phenomenon: disk winds.
Protoplanetary disks are thought to inherit magnetic flux from their parent molecular clouds. Depending on the strength and inclination of the threading magnetic fields, the centrifugal force along the field lines or the magnetic field pressure in the vertical direction can accelerate the material on the disk surface to the escape velocity \citep{Blandford1982,Shibata1986}. The magnetically driven winds also induce gas accretion {\it within the disks} because the magnetically accelerated wind material takes away the disks' angular momentum. Recent MHD simulations have shown that the wind-driven accretion alone can account for the observed accretion rates for protoplanetary disks if the magnetic fields vertically threading the disks are sufficiently strong \citep{BaiStone2013,Bai2013,Simon2013a,Lesur2014,Gressel2015}.
However, lacking direct measurements of the strength of disks'  large-scale magnetic fields (although there are some constraints on Solar nebula magnetic field strengths from paleomagnetic measurements of meteorites; see \citealt{Fu2014,Wang2017}), it is unclear whether realistic protoplanetary disks can retain the required amount of the vertical magnetic flux. There are several recent theoretical studies in this direction \citep{Guilet2014,Okuzumi2014,Takeuchi2014,BaiStone2017,Zhu2018,Leung2019}.

There are purely hydrodynamical (i.e, non-MHD) instabilities that can produce weak turbulence, including the vertical shear instability \citep{Urpin1998,Nelson2013}, convective overstability \citep{Klahr2014,Lyra2014}, and zombie vortex instability \citep{Marcus2015}. These hydrodynamical instabilities operate in different ranges of the gas cooling timescale and hence in different disk regions, with the vertical shear instability being most relevant to cold outer disk regions where comets form \citep{Malygin2017,Lyra2019}.  According to hydrodynamical simulations, turbulence driven by these instabilities does transfer the disk's angular momentum radially, but its transport efficiency appears to be low compared to strong MRI-driven turbulence and magnetically driven winds \citep{Lyra2019}. This does not mean that hydrodynamical instabilities are negligible because they can play a significant role in the transport of disk material. For instance, the vertical shear instability produces vertically elongated turbulent eddies that strongly diffuse gas and dust in the vertical direction \citep{Flock2020}. The hydrodynamical instabilities also cause the formation of gas pressure bumps and vortices that can efficiently trap dust particles \citep{Flock2020,Raettig2021}. 

Massive planets also have interesting effects on disk gas dynamics. Early theoretical studies already predicted that a planet larger than Neptune can carve an annular density deficit called a gap in the background gas disk \citep{Lin1993}. Recent three-dimensional simulations have shown that a massive planet also induces meridional (i.e., radial and vertical) flows of the disk gas both inside and outside the gap \citep{Morbidelli2014,Fung2016} (Fig. \ref{fig:schematic}). Signatures of the meridional flow are obtained from a detailed analysis of high-resolution molecular line data \citep{Teague2019}. The gaps and meridional flows induced by massive planets have important implications for dust evolution (see Sect.~2.3).  

\subsection{Temperature distribution and snowlines}
\label{sec:temperature}

As in the present-day Solar system, the temperature in protoplanetary disks generally decreases with orbital radius. Therefore, each chemical species in a disk has a critical radius inside which its solid form is unstable.
For volatile species, these critical radii are called the snowlines. Because different volatile species have different sublimation temperatures, each of them has its own snowline. [Considering the temperature gradient in the vertical direction, the boundary is actually a snow surface. The snowline is an intersection between the snowsurface and the disk midplane as depicted in Fig 1.  We focus on the snowline here, because the midplane dominates in the mass distribution.]

The snowlines are generally expected to affect the radial distribution of volatiles in solids and also in the gas \citep{Hayashi1981,Oberg2011}. If we fully understand where the snowlines are and how they move with time, we will be able to constrain where and when planets and comets of different compositions form in disks. 

\begin{figure}[t]
\centering
\resizebox{\hsize}{!}{\includegraphics[bb=0 0 288 263]{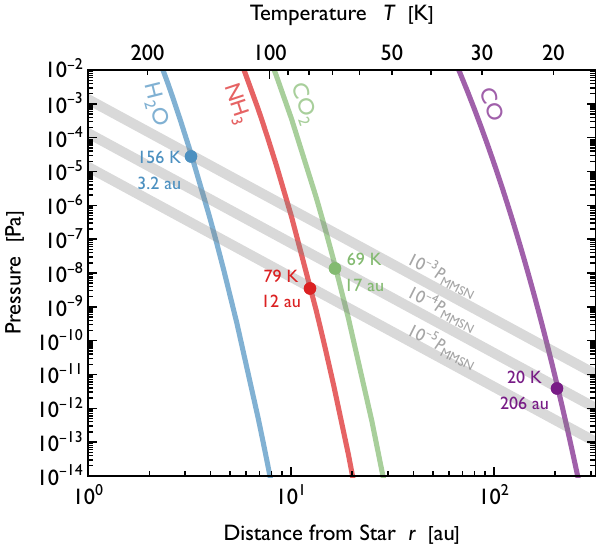}}
\caption{Saturation vapor pressures for pure ${\rm CO}$, ${\rm CO}_2$, ${\rm NH}_3$, and ${\rm H_2O}$ ices (thin lines) as a function of temperature $T$ (indicated on the top edge). For reference, the thick lines show how the temperature and 0.1, 0.01, and 0.001\% of the total midplane gas pressure vary with orbital radius (indicated on the bottom edge) in a radially extended version of the classical, optically thin minimum-mass Solar nebula \citep{Hayashi1981}. The circles mark the locations of the snowlines.
}
\label{fig:Pev}
\end{figure}

Themodynamically, the snowline of each volatile species can be defined as the location where its saturation vapor pressure equals its partial pressure in the gas phase. To illustrate how the locations of the snowlines are determined, we plot in Fig. \ref{fig:Pev} the saturation vapor pressure curves for pure ${\rm CO}$, ${\rm CO}_2$, ${\rm NH}_3$, and ${\rm H_2O}$ ices as a function of radial distance $r$ for the classical, optically thin minimum-mass Solar nebula (MMSN) model of \citet{Hayashi1981}, which has radial temperature and midplane pressure profiles of $T_{\rm MMSN} = 280(r/1{~\rm au})^{-1/2}~\rm K$ and $P_{\rm MMSN} = 1.4(r/1{~\rm au})^{-13/4}~\rm Pa$, respectively. 
Here, we adopt analytic expressions for the saturation vapor pressures as a function of temperature provided by \citet{Bauer1997} (for H$_2$O) and by \citet{Yamamoto1983} (for the other species), and plot them as a function of $r$ using the assumed temperature profile.
Let us assume that the disk contains ${\rm CO}$, ${\rm CO_2}$, ${\rm NH_3}$, and ${\rm H_2O}$ at spatially uniform molar abundances of $f = 10^{-4}$, $10^{-4}$, $10^{-5}$, and $10^{-3}$, respectively, so that the relative abundances between the four volatiles are crudely consistent with the abundances in comets \citep{Mumma2011}.  If all the volatiles were in the gas phase, they would have partial pressures of $f P_{\rm MMSN}$. By comparing the saturation vapor pressures and assumed partial pressures, we find that the snowlines of ${\rm CO}$, ${\rm CO_2}$, ${\rm NH_3}$, and ${\rm H_2O}$ are located at $r \approx$ 206, 17, 12, and 3.2 au, with sublimation temperatures of $\approx 20$, $69$, $79$, and $156$ K, respectively, in this particular disk model (see the circles in Fig. \ref{fig:Pev}).
[The abundance of H$_2$O relative to H$_2$ is set to $10^{-3}$ assuming that H$_2$O is the dominant reservoir of oxygen \citep{Loddard03}. If water is mostly inherited from molecular clouds, H$_2$O abundance would be $10^{-4}$ \citep{Whittet93}. Then the water snowline is slightly shifted outwards.  We also note that the ices would be a mixture of various molecules, while we consider pure ices here for simplicity. Sublimation of mixed ice is more complicated than that of pure ice  \citep[e.g. see review by ][]{Hama2013, minissale2022}. For example, laboratory experiments show that the adsorption energy and thus the sublimation temperature of a molecule depends on the composition and surface structure (e.g. crystal or amorphous) of the substrate ice. Referring to the observations of comets and interstellar ice, H$_2$O is expected to be a dominant composition of ice in disks. Then a fraction of molecule with higher volatility, such as CO and CO$_2$, would be trapped in H$_2$O ice and desorb at H$_2$O snow line.]

The strong dependence of the saturation vapor pressures on temperature illustrated in Fig. \ref{fig:Pev} clearly indicates that realistic modeling of the disk temperature structure is key to accurately infer the snowline locations. Unfortunately, the simple optically thin disk model adopted above does not apply to protoplanetary disks, which are mostly optically thick to radiation from the central star; in the MMSN, for instance, an optical thickness to stellar radiation can be as high as $\sim 10^5$. Optically thick disks can receive stellar radiation only at their surfaces, and hence their interior temperatures tend to be lower than in optically thin disks unless any internal heat source is present. 
Specifically, for an optically thick disk passively heated by a central star of luminosity $L_*$ and mass $M_*$, the temperature of dust grains at the optically thin surface and optically thick midplane can be estimated as \citep{Kusaka1970,Chiang1997}
\begin{equation}
T_{\rm irr,surf} \approx 490\pfrac{\epsilon}{0.1}^{-1/4} \pfrac{L_*}{L_\odot}^{1/4}\pfrac{r}{1~\rm au}^{-1/2}~\rm K,
\label{eq:T_irr_surf}
\end{equation}
\begin{equation}
T_{\rm irr,mid} \approx 120 \pfrac{L_*}{L_\odot}^{2/7}\pfrac{M_*}{M_\odot}^{1/7}\pfrac{r}{1~\rm au}^{-3/7}~\rm K,
\label{eq:T_irr_mid}
\end{equation}
respectively,
where $\epsilon$ is the ratio of the infrared to visible absorption cross sections of the grains at the disk surface.
The temperature profile $T_{\rm MMSN}$ of the MMSN disk model is equivalent to $T_{\rm irr, surf}$ with $\epsilon = 1$ and $L_* = 1 L_\odot$, where the value $\epsilon = 1$ applies to dust particles larger than microns. We here assume that only submicron-sized grains remain in the surface region and adopt $\epsilon \sim 0.1$ \citep{Isella2005}. 
The estimate for $T_{\rm irr, mid}$ depends weakly on the height of the surface where the starlight is finally absorbed \citep{Chiang1997}. The scale height, and thus $T_{\rm irr, mid}$ depends on the mass of the central star. In Equation~\eqref{eq:T_irr_mid}, we have assumed that the starlight absorption surface lies at three scale heights above the midplane. 

\begin{figure*}[t]
\centering
\resizebox{0.45\hsize}{!}{\includegraphics[bb=0 0 288 473]{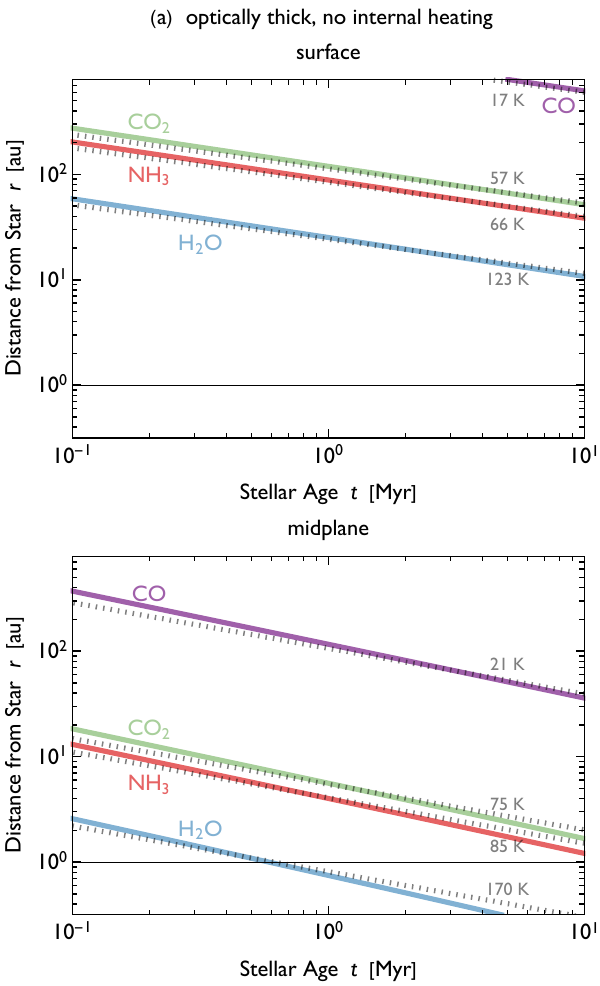}}
\hspace{2mm}
\resizebox{0.45\hsize}{!}{\includegraphics[bb=0 0 288 473]{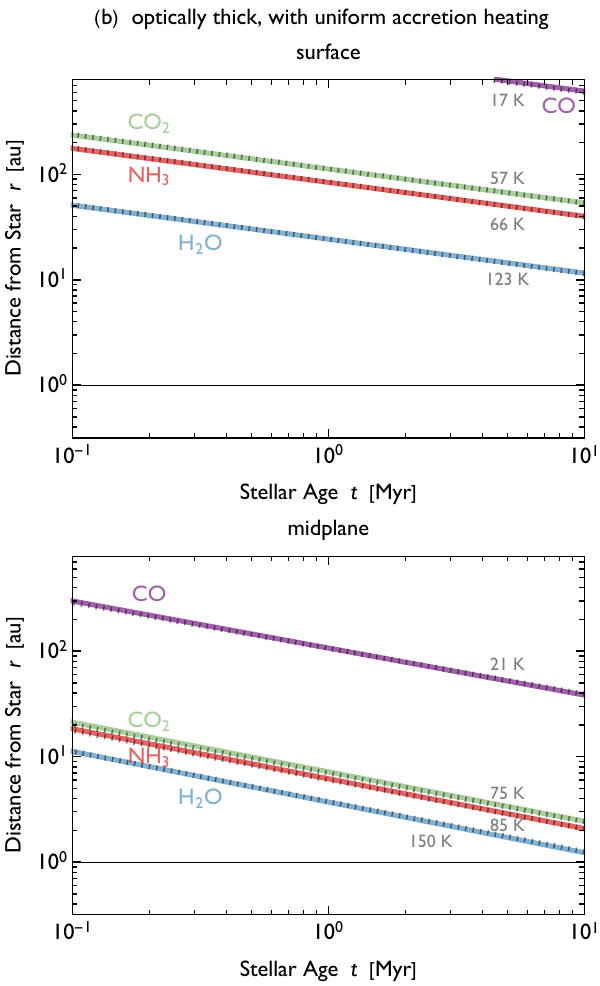}}
\caption{ 
Locations of the snowlines for pure CO, CO$_2$, NH$_3$, and H$_2$O ice at the optically thin surface (upper panels) and at the optically thick midplane (lower panels) as a function of stellar age $t$ for disks around a solar-mass pre-main-sequence star. The molar abundances of ${\rm CO}$, ${\rm CO_2}$, ${\rm NH_3}$, and ${\rm H_2O}$ are assumed to be $10^{-4}$, $10^{-4}$, $10^{-5}$, and $10^{-3}$, respectively, everywhere in the disks. Panels (a) are for an optically thick MMSN heated by stellar radiation only, with the snowlines migrating as the stellar luminosity evolves. Panels (b) are for an optically thick, viscous accretion disk with uniform accretion heating in addition to stellar irradiation. In this latter model, the evolution of both the stellar luminosity and accretion rate causes the migration of the snowlines.
The dotted lines are contours of constant temperatures, and the thin horizontal line marks the location of the current Earth's orbit, $r = $ 1 au. 
}
\label{fig:Tmid}
\end{figure*}
The large temperature difference between the midplane and surface regions of the disks has significant implications for the compositional distribution of solids in the disks.
Fig. \ref{fig:Tmid}(a) illustrates how the locations of the snowlines of some major volatiles at the surface (here referring to three scale heights above the midplane) and at the midplane evolve in an optically thick, passively irradiated disk (see also the snow surface in Fig. \ref{fig:schematic}). Here, the central star is assumed to have a fixed mass of $M_* = 1M_\odot$, while the stellar luminosity is assumed to decrease from $10 L_\odot$ at $t = 0.1~{\rm Myr}$ to $0.5 L_\odot$ at $t = 10~\rm Myr$ in a power-law fashion, mimicking the evolution of solar-mass pre-main-sequence stars. For simplicity, the abundances of the volatiles are assumed to be spatially uniform and equal to the values used in Fig. \ref{fig:Pev}, and the surface density profile of the MMSN is used to calculate the disk pressure distribution.
We find from the lower panel of Fig. \ref{fig:Tmid} (a) that the H$_2$O snowline at the midplane would move inside 1 au---the radius of the current Earth's orbit---within the first million years of star and planet formation. In contrast, the H$_2$O snowline in the optically thin surface region lies farther away, at more than $10$ au from the central star (see the upper panel of Fig. \ref{fig:Tmid} (a)). This difference arises mainly from the vertical temperature (rather than pressure) gradient.

The midplane temperature given by Equation~\eqref{eq:T_irr_mid} should be taken as a lower limit because we have neglected any internal heat sources.
The classical accretion disk model that assumes vertically uniform turbulent viscosity \citep{Lynden-Bell1974} predicts that the gravitational energy liberated by mass accretion toward the star can significantly heat the disk interior, in particular in the early disk evolutionary phase where both the accretion rate and disk surface density are high. Fig. \ref{fig:Tmid} (b) illustrates the evolution of the midplane temperature for a disk heated by internal viscous heating in addition to stellar radiation. We consider a solar-mass star and let its luminosity evolve as in Fig.~\ref{fig:Tmid} (a). The disk accretion rate is assumed to decay as $\dot{M} = 4\times 10^{-8} (t/1~{\rm Myr})^{-1.07}~M_\odot~\rm yr^{-1}$ \citep{Hartmann2016}. The disk opacity is taken to be $2(T/100~\rm K)^2~\rm cm^2~\rm g^{-1}$ \citep{Bell1994}, and the disk's turbulent viscosity is taken to be 1 \% of the local sound velocity times the gas scale height. 
The figure shows that the viscous heating pushes the H$_2$O snowline out beyond 1 au during the entire  Class II phase ($t \lesssim $ several Myr). 
Note that the temperature evolution illustrated here depends on the disk opacity assumed, and hence on the evolution of dust grains that dominate the opacity.
In the disk model adopted here, a factor of 10 decrease in the opacity leads to a factor of $\approx$ two decrease in the midplane temperature. 

However, as already mentioned in Sect.~\ref{sec:gas}, theoretical studies call into question the applicability of the simple viscous accretion disk model. A more realistic picture of disk accretion may be that the accretion is mainly driven by magnetic fields. In this picture, disk heating still occurs through the Joule dissipation of electric currents associated with the magnetic fields. However, the Joule heating tends to predominantly occur near the disk surface because the low electric conductivity around the midplane prohibits the development of strong currents. Thermal radiation generated near the surface region can easily escape and is therefore inefficient at heating the midplane region. For this reason, the temperature structure of magnetically accreting disks tends to be close to that of passively irradiated disks as shown in Fig. \ref{fig:Tmid}(a) unless both the disk opacity and ionization fraction are sufficiently high \citep{Hirose2011,Bethune2020,Mori2019,Mori2021}.

The disk temperature structure and the locations of the snowlines drastically change during episodic accretion outburst events introduced in Sect.~\ref{sec:gas}. For example, V883 Ori is a solar-mass FU Ori star with an elevated luminosity of $\sim 200 L_\odot$ \citep{Furlan2016} and a massive circumstellar disk.
Equation~\eqref{eq:T_irr_mid} implies that the midplane of the V883 Ori disk in the outburst phase should be $\sim 4.5$ times hotter than those of disks around solar-luminosity stars.
Assuming a sublimation temperature of 150 K for H$_2$O ice, we find that the strong irradiation heating should push the H$_2$O snowline out to $r \sim $ 20 au. In reality, the bolometric luminosity of $\sim 200 L_{\odot}$ includes accretion heating, which could push the snowline further out by heating the disk from inside.

These theoretical estimates of disk temperature structure are compared with SED and spatially-resolved observations of dust continuum and molecular lines. \citet{Cieza2016} observed 1.3 mm dust continuum towards V883 Ori to find that the brightness and optical depth sharply rise inward at $\sim 40$ au. A plausible interpretation is that there is a water snowline at 40 au and ice sublimation causes the abrupt change of the dust grain size and opacity  \citep[see also section 4.2 and][]{schoonenberg2017}. More recently, \citet{Law21} derived the 2D (radial and vertical) gas temperature distribution of five disks using high- resolution observations ($\sim 20$ au) of CO and its optically thick isotopologues. The derived temperatures are roughly in agreement with the simple estimates presented in Fig, \ref{fig:Tmid} \citep[see also][]{Dutrey2017,Flores2021}.

\subsection{Dust growth: theory} \label{sec:dust_growth}
In studies of planet formation and protoplanetary disks, dust can refer to any solid particles smaller than planetesimals, not only  (sub)micron-sized dust grains but also pebble to boulder-sized solid particles. Although the solids only comprise a minor fraction of the mass in disks, they play significant roles in planet formation and disk evolution. The solids are the ultimate building blocks of all solid bodies in planetary systems including comets. Small dust grains govern the disks' thermal structure, and their size distribution can affect snowline locations \citep{Oka2011} and the disks' hydrodynamical instabilities \citep{Malygin2017,Barranco18,Fukuhara2021}. Thermal emission from micron to millimeter-sized dust particles dominates the disks' continuum emission that we observe with infrared and radio telescopes. Small grains can even regulate the disk MHD by facilitating the recombination of ionized gas particles interacting with magnetic fields \citep{Sano2000,Ilgner2006,Wardle2007,Bai2011}.
Therefore, to understand how solid bodies form in protoplanetary disks, we must fully understand how dust evolves and how it interacts with the disk gas. 

From the theoretical point of view, there is no doubt that dust growth starts already in the earliest stage of disk evolution, at least in inner disk regions. Models of dust evolution in both laminar and turbulent disks \citep{Weidenscilling1980,Nakagawa1981,Dullemond05,Tanaka05,Brauer2008,Birnstiel2010} show that micron-sized grains grow into 0.1--1 mm-sized grains on a timescale of $\ll 1~\rm Myr$.
In moderately turbulent disks, a useful simple estimate for the local growth timescale (size-doubling time) is available \citep{Takeuchi2005,Brauer2008}: 
\begin{align}
    t_{\rm grow} &\sim \frac{\Sigma_{\rm gas}}{\Sigma_{\rm dust}}\frac{1}{\Omega_{\rm K}} 
    \notag \\
    &\sim 20\pfrac{\Sigma_{\rm gas}/\Sigma_{\rm dust}}{100}\pfrac{r}{1~\rm au}^{3/2}\pfrac{M_*}{M_\odot}^{-1/2}~\rm yr,~~
    \label{eq:tgrow}
\end{align}
where $\Sigma_{\rm gas}$ and $\Sigma_{\rm dust}$ are the mass surface densities of gas and dust, respectively, and $\Omega_{\rm K}$ is the local Keplerian frequency.
Equation~\eqref{eq:tgrow} does not involve turbulence strength because the negative effect of turbulence on dust settling cancels its positive effect on particle collision speeds. Furthermore, Equation~\eqref{eq:tgrow} is independent of particle size, implying that the particle radius increases exponentially with time. For instance, micron-sized grains at 30 au would grow to millimeter-sized aggregates in $\sim 2\times 10^4~\rm yr$. 
Strictly speaking, Equation~\eqref{eq:tgrow} overestimates the growth time of small grains that are vertically well mixed in the disk, so the timescale on which millimeter-sized aggregates form can be even shorter. Highly porous (or ``fluffy'') dust aggregates can also grow faster than estimated by Equation~\eqref{eq:tgrow} \citep{Okuzumi12,Kataoka13,Garcia20}.
In any case, the growth timescale at 30 au is much shorter than the typical ages of Class II disks ($\sim 10^6~\rm yr$) and is even comparable to the ages of the youngest Class 0 sources ($\sim 10^4~\rm yr$). 

Equation~\eqref{eq:tgrow} assumes that grain collisions always result in sticking. This assumption is valid for micron-sized grains with low collisional velocities, unless they are strongly negatively charged in the partially ionized protoplanetary disks \citep{Okuzumi2009}.
However, as the grain aggregates grow, they obtain a higher collision velocity induced by turbulence \citep{Ormel07} and by radial drift on the background gas \citep{Whipple1972,Adachi1976,Weidenschilling1977}.
The increased collision velocity may prevent further growth by inducing collisional fragmentation \citep{Blum2000,Guettler2010} and bouncing \citep{Guettler2010,Zsom2010}. Moreover, in a smooth gas disk with a negative radial pressure gradient, their radial drift is inward (see the chapter by Simon et al.).
For millimeter and centimeter-sized aggregates, the timescale of the radial inward drift can be shorter than the local growth timescale $t_{\rm grow}$, meaning that these pebbles would fall toward the central star before growing to meter-size boulders. 
Therefore, a more realistic picture of dust growth is that grains rapidly grow initially and then reach the maximum size set by fragmentation, bouncing, or radial drift  \citep{Birnstiel12,Drazkowska23}. These growth barriers complicate planetesimal formation  (see also the chapter by Simon et al.). 

In the context of this chapter, it is important to point out that grains' chemical composition is key to understanding how they grow. Early models and experiments suggested that aggregates made of water ice are sticker than silicates \citep{Dominik97} and may even overcome the fragmentation barrier \citep{Wada09}. Such sticky grains also tend to form highly porous aggregates, which is beneficial for overcoming the drift barrier because the porous aggregates grow rapidly  \citep{Okuzumi12,Kataoka13}. However, this sticky water ice scenario is questioned by recent experiments showing that water ice is not so sticky at low temperatures \citep{Gundlach18,Musiolik19}. Instead, some recent studies suggest that silicates are sticker than previously thought \citep{Kimura15,Steinpilz19}. Some (if not all) types of organic matter are also sticky in a warm environment \citep{Kouchi02,Piani17,Bischoff20}. CO$_2$ ice appears to be less sticky than water ice \citep{Musiolik16a,Musiolik16b,Arakawa21,Fritscher21}. These imply that the fate of dust coagulation and planetesimal formation may depend on temperature and hence on distance from the central star \citep{Birnstiel2010,Pinilla17,Homma19,Okuzumi19}.

Ice sublimation, condensation, and sintering around snowlines can produce local pileups of solids \citep{Cuzzi04,Saito11,Sirono11,Okuzumi16,SchoonenbergOrmel17,Drazkowska17,Ida16,Ida21,Hyodo19,Hyodo21}. These pileups can have important implications for planetesimal formation and the disks' observational appearance (see also Section~\ref{sec:substructures}).

\subsection{Dust growth: observations}\label{sec:dust_growth_obs}

Dust grains are the dominant source of disk opacities. Continuum emission and scattering thus provide information on the physical properties of grains (including size, porosity, and composition). Here, we briefly review important constraints on the degree of dust growth in disks obtained from radio observations. For more comprehensive reviews on disk dust observations, we refer to \citet{Testi14} and \citet{Miotello23}.

The spectral index of dust continuum emission in the millimeter wavelength has been used to constrain the maximum grain size in disks; it is defined by
\begin{equation}
    \alpha_\nu = \frac{d\ln I_\nu}{d\ln \nu},
\end{equation}
where $I_\nu$ is the intensity of the emergent radiation at frequency $\nu$. For a black body of uniform temperature $T$, $I_\nu$ is equal to the Planck function $B_\nu(T)$ for the temperature, and one has $\alpha_\nu = 2$ in the Rayleigh--Jeans limit ($B_\nu \propto \nu^2$). For an optically thin disk, one has $I_\nu \approx \kappa_\nu \Sigma_{\rm dust}B_\nu$, where $\kappa_\nu$ is the dust absorption cross section per unit dust mass \citep{Miyake1993}. In the latter case, $\alpha_\nu$ reflects the frequency dependence of $\kappa_\nu$, which in turn reflects the size distribution of opacity-dominating dust grains. 
For example, interstellar dust grains of maximum size $\sim 0.1~{\rm \mu m}$ yield $\kappa_{\nu} \propto \nu^2$ and hence $\alpha_{\nu} \approx 4$ at radio wavelengths. As the grains grow and their maximum grain size exceeds the wavelengths, the $\nu$ dependence of $\kappa_{\nu}$ becomes weaker, leading to $\alpha_{\nu}< 4$ \citep[e.g.,][]{Miyake1993}.
Constraining the grain size from $\alpha_\nu$ is also possible for disks that are optically thick but have a non-zero albedo $\omega_\nu$, for which $I_\nu \sim \sqrt{1-\omega_\nu} B_\nu$ \citep{Rybicki1979}. The factor $\sqrt{1-\omega_\nu} (<1)$ represents the effect of multiple scattering by dust particles in an optically thick disk suppressing the disk's thermal emission \citep{Rybicki1979,Miyake1993,Birnstiel2018,Liu2019,Zhu2019,Sierra2020}.

Early millimeter and submillimeter surveys \citep{Weintraub1989,Beckwith1990,Beckwith1991,Andrews2005,Andrews2007,Ricci2010a,Ricci2010b} already showed that the spatially integrated continuum emission from T Tauri disks has spectral indices of $\sim 2$--3 in the (sub)millimeter range. 
More recently, high-resolution millimeter interferometric observations have provided information on the spatial variation of $\alpha_\nu$ in individual disks \citep{Perez2012,Perez2015,ALMA2015,Guidi2016,Tazzari2016,Huang2018,Cazzoletti2018,Dent2019,Carrasco2019,Soon2019,Macias2021}. Overall, local values of $\alpha_\nu$ range between 1.5--4 and appear to be anticorrelated with millimeter intensity. This anticorrelation indicates either that particles larger than a millimeter are concentrated in the brighter regions or that the brighter regions simply have higher optical depths. In some disks, the spectral index approaches $\sim 4$ toward the disks' outer edges  \citep{Perez2012,Guidi2016,Tazzari2016,Dent2019,Carrasco2019}, suggesting that the grains in these outermost regions are smaller than millimeter in size \citep{Miyake1993}.
The values of $\alpha_\nu \la 2$ can be explained by optically thick emission from dust grains that have higher albedos at shorter wavelengths. In particular, in the (sub)millimeter wavelength range, $\alpha_\nu \la 2$ implies a population of dust grains with maximum grain sizes of 0.1--1 mm \citep{Liu2019,Zhu2019,Sierra2020}.  
An alternative explanation is that the disks with $\alpha_\nu < 2$ have temperatures that decrease with distance from the midplane \citep{Sierra2020}.

If we assume that the entire disk is optically thin at the observed wavelengths, the spectral index of $\sim 2$--3 indicates grain growth to millimeters in radius \citep{Miyake1993,Draine2006}.
However, it is not evident that the disks, even its outer part, are always optically thin at submillimeter and millimeter wavelengths. Inferring disks' optical thicknesses is intrinsically difficult because the disks' temperature distribution is unknown a priori. \citet{Huang2018} and \citet{Dullemond2018} used a simple model for the temperature of passively irradiated disks (similar to Equation~\eqref{eq:T_irr_mid}) and found that some bright dust rings observed in ALMA millimeter images have similar millimeter absorption optical thicknesses of $0.2$--1. Although one interpretation is that the rings are barely optically thin by coincidence,  another possibility is that the rings are actually optically thick but appear to be darker than the black body because of scattering, $I_\nu \sim \sqrt{1-\omega_\nu} B_\nu$ \citep{Zhu2019}.

Recently, radio polarimetric observations have been used to derive independent constraints on the grain size in the outer regions ($\gtrsim$ 10 au) of protoplanetary disks. The observations have shown that many inclined disks produce uniformly polarized submillimeter continuum emission whose polarization direction is parallel to the disks' minor axes \citep[e.g.,][]{Stephens2014,Stephens2017,Hull2018,Harris2018,Cox2018,Sadavoy2018,Dent2019}. This polarization pattern can be explained if the polarized continuum emission is dust thermal radiation scattered by the dust particles themselves \citep{Kataoka2015,Yang2016}. In this interpretation, the observed degree of polarization gives a strong constraint on the size of the opacity-dominating dust particles because the scattering and polarization efficiencies depend strongly on grain size. Models assuming spherical, compact silicate particles \citep{Kataoka2015,Yang2016} predict a maximum grain size of $\sim 100~\rm \mu m$ for the disks with uniformly polarized submillimeter emission. 

This new constraint has garnered considerable attention in recent years because the inferred grain size is an order of magnitude smaller than estimated from the spectral index. 
The cause of the discrepancy between the two grain size estimates is under debate. In fact, the estimates from disk polarized emission largely depend on the grain properties assumed, including grain shape \citep{Kirchschlager2020}, porosity \citep{Tazaki2019,Brunngraber2021}, and composition \citep{Yang2020}. 
For disks that are optically thick at submillimeter wavelengths, the discrepancy may be resolved if grains larger than $\sim 100~\rm \mu m$ have already settled to the optically thick midplane region and give no contribution to submillimeter emission, while still affecting the spectral index by contributing to millimeter emission that is optically thinner \citep{Brunngraber2020,Ueda2021}. 

\subsection{Dust spatial distribution}\label{sec:substructures}
With ALMA, we can now observe dust emission from protoplanetary disks in nearby star-forming regions at a spatial resolution of $\sim 5$ au. Arguably the most striking discovery from the ALMA disk observations is the prevalence of small-scale structures---rings, gaps, spirals, and crescents---in dust thermal emission \citep[e.g.,][]{ALMA2015,Andrews2018,Long2018,Cieza2021}.
Substructures on similar or smaller scales are also found in near-infrared scattered light images that probe the disk surfaces \citep[e.g.,][]{Avenhaus18,Garufi18}. 
We refer to \citet{Andrews2020} for a review of recent disk substructure observations at both radio and infrared wavelengths. 

\begin{figure*}[t]
\centering
\resizebox{0.75\hsize}{!}{\includegraphics[bb=0 0 1870 620]{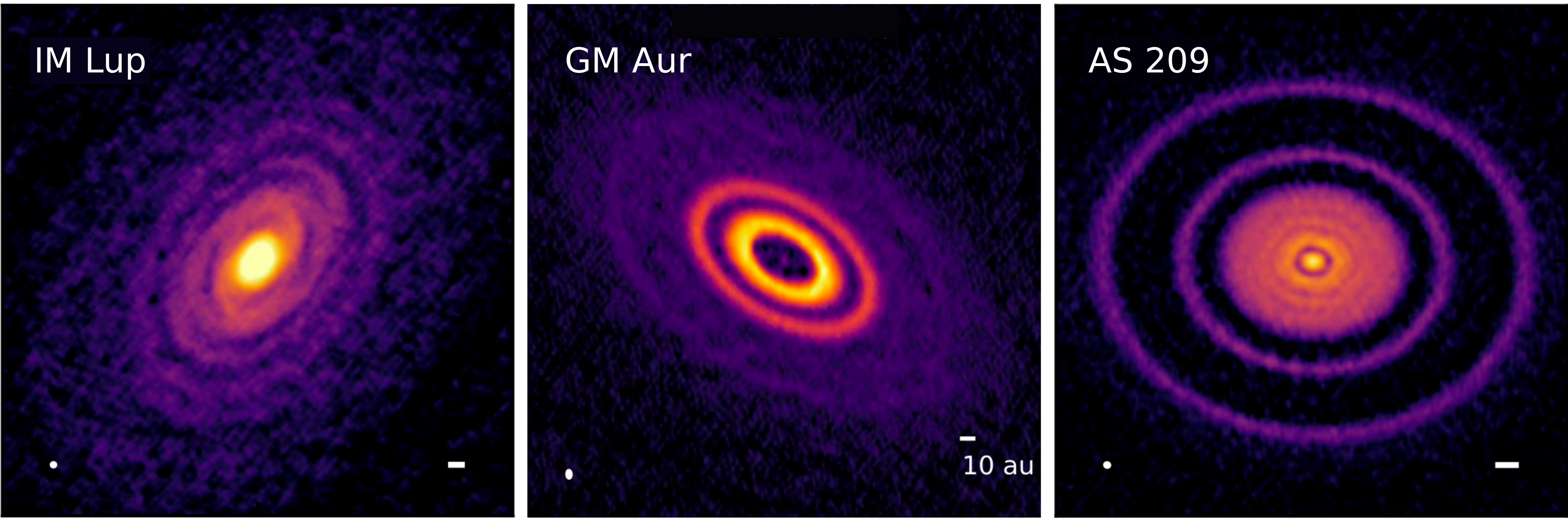}}
\caption{
ALMA millimeter continuum images of three selected protoplanetary disks (IM Lup and AS 209 adapted from \citealt{Andrews2018}; GM Aur from \citealt{Huang2020}) showing ring/gap substructures in the disks' dust component. Beam sizes and 10 au scale bars are shown in the bottom left and right corners of each panel, respectively. The image of the IM Lup disk also exhibits spirals.
}
\label{fig:substructures}
\end{figure*}

The most commonly observed substructures are axisymmetric rings and gaps. They were first discovered in the disk around HL Tau \citep{ALMA2015} and have since been found in many large, bright Class II disks \citep{Huang2018,Long2018}. Fig. \ref{fig:substructures} shows ALMA dust continuum images of the well-studied protoplanetary disks around IM Lup, GM Aur, and AS 209. It illustrates the diversity of the ring/gap substructures: whereas the disk around IM Lup only exhibits shallow gaps (plus spirals), the AS 209 disk features deeper, more extended gaps together with remarkably narrow rings. The disk around GM Aur has a ring connected to a fainter outer disk, a feature also visible in the dust images of some other disks (e.g., Elias 24; \citealt{Huang2018}).

The origins of these dust substructures are a subject of active research \citep{Andrews2020}.
A widely accepted explanation for narrow dust rings is dust trapping in ring-shape pressure maxima (bumps) \citep{Pinilla12,Dullemond2018,Rosotti2020}, where the radial inward drift of dust particles halts \citep{Whipple1972}. This scenario is tempting because dust trapping at pressure bumps has long been anticipated as a solution to the radial drift problem in planetesimal formation \citep{Johansen2014}.  
There are a variety of physical mechanisms potentially yielding annular pressure bumps, including planet--disk interaction \citep{Paardekooper2006}, MHD effects \citep{Johansen2009,Suriano2018,Flock2015}, and dust--gas interaction \citep{Gonzales2018}. Planets can also produce gaps and spirals.

In principle, snowlines could also produce dust rings and gaps as mentioned in Section~\ref{sec:dust_growth}. \citet{Zhang15} and \citet{Okuzumi16} argued that the snowlines of major volatiles, including H$_2$O and CO, are responsible for the major dust rings and gaps in the HL Tau disk. However, the current understanding is that most of the rings and gaps that have been observed in a number of disks are not relevant to snowlines \citep{Huang2018,Long2018,vanderMarel19}. The main reason is that the radial positions of the substructures do not appear to be correlated with the central star's luminosity, which should determine snowline locations. Nevertheless, there are a few disks (HD 163296 and MWC 480 in addition to HL Tau) in which a dust ring is found close to the expected location of the CO snowline \citep{zhang21}.

ALMA observations of dust emission have also provided important constraints on the dust surface density in the outer ($r \gtrsim~10~\rm au$) disk region where icy bodies like comets form. If we assume that this region is optically thin to its own millimeter emission, the dust surface density is derived from the millimeter intensity $I_\nu$ as $\Sigma_{\rm dust} = I_\nu/(\kappa_\nu B_\nu)$ (see also Section~\ref{sec:dust_growth_obs}).
For example, the dust surface densities of the IM Lup, GM Aur, and AS 209 disks are estimated to be comparable or an order of magnitude higher than that in the MMSN
($\Sigma$(ice+rock)$= 30(r/1{\rm au})^{-3/2}$ g cm$^{-2}$)
for a reasonable assumption about $\kappa_\nu$ for 0.1--1 mm sized grains (see the bottom panels of Fig. \ref{fig:radial_dist} in Section~3). We note, however, that these particular objects are among the largest, most massive disks; their dust surface densities may not represent those of more common, compact disks \citep{Miotello23}.

\section{Chemistry in Protplanetary disks} \label{s:sec3}

Comets contain a significant amount of icy material, which has been investigated to reveal the thermal history and chemical environment in the cold outer regions of the Solar nebula (e.g. chapter by Biver et al.). For example, the composition of major cometary ice is often similar to that of interstellar ice. The relative abundance of CO$_2$ and CH$_3$OH to water is $2-30 \%$ and $0.2-7 \%$ in comets, while these ratios are $15-44 \%$ and $5-12 \%$ in quiescent molecular clouds \citep{Mumma2011}. Inheritance of interstellar ice in comets has thus long been discussed.
Observations of protoplanetary disks, on the other hand, can reveal the spatial distribution and evolution of molecules at the sites of ongoing planetary-system formation.
These two kinds of information are complementary to elucidate the formation of comets and the Solar system.

Theoretical and observational studies show that various chemical processes are going on within the disks \citep[e.g.][]{Aikawa2002,semenov2013,oberg23}.
As we have seen in the previous section, there are large gradients of temperature and density in both the radial and vertical directions in a disk (Fig. \ref{fig:schematic}). In the vertical directions, gases are in hydrostatic equilibrium and the density decreases towards the disk surface. The disk surface is directly irradiated by the stellar radiation and thus is warm. Molecules are dissociated to radicals and atoms by UV radiation from the central star \footnote{While the surface temperature of T Tauri stars is $\sim 4000$ K, UV radiation is produced by the gas accretion from the disk to the star.} and interstellar radiation field; the disk surface is a photon-dominated region (PDR). Beneath the surface layer, there is a warm molecular layer ($\gtrsim 20$ K), which is moderately shielded from UV radiation and thus harbors various molecular gas such as CO. Ion-molecule reactions are triggered by X-rays from the central object \footnote{X-rays are emitted due to magnetic activity on the stellar surface.} and cosmic rays. In the midplane, the density is highest and dust grains could be further concentrated due to sedimentation. Since the temperature decreases with radial distance, snowlines are defined in the midplane as explained in \S 2.2 (Fig. \ref{fig:Pev}). Thermal energy could be high enough to drive chemical reactions in the inner hot regions \citep[e.g. above several hundreds of K at $< 1$ au,][]{Yang2013}.
Even in the outer cold midplane, reactions in the gas phase and within ices can be triggered by high-energy cosmic rays, X-rays, and decay of radioactive nuclei (e.g. $^{26}$Al).
Such layered structures, i.e. the PDR layer, warm molecular layer, and icy cold midplane, are confirmed by spatially and spectrally resolved observations (i.e. channel maps) of disks \citep{Dutrey2017,Ruiz-Rodrigurz_2021,Rosenfeld2013,Law21}. While the chemical composition of ice in the midplane would be most directly relevant to comets, observation of such ice is not easy (see \S 3.3). Gaseous line observations in millimeter and submillimeter wavelengths are much more sensitive to low-abundance species. Chemistry in the gas and ice is linked via freeze-out, sublimation, and radial/vertical transport as we will discuss below.  

In this section, we overview observations of cold molecular gas ($\lesssim 100$ K) in millimeter and submillimeter wavelengths (\S 3.1) and warm molecular gas ($\gtrsim 100$ K) and ice in infrared and shorter wavelengths (\S 3.2 and \S 3.3; see also Fig. \ref{fig:schematic}).
Table \ref{tab:molecules} lists the molecular species so far detected in Class II disks in the order of increasing number of atoms and molecular mass. The observed wavelength is listed as well, since it tells where these molecules reside, i.e. cold or warm regions.
In addition to the molecular abundances, we cover isotope fractionation and spin temperatures in section \S 3.4.

\begin{table*}[h]
\caption{Molecules detected in Class II disks\tablenotemark{a}}
\label{tab:molecules}
\begin{small}
 \begin{tabular}{ccc  ccc }
 \hline
species & wavelength & references (e.g.) & species & wavelength & references (e.g.)\\
\hline
H$_2$  & NIR and MIR  & [1][2]  & HD     &  FIR & [3][4]\\
CH$^+$ & FIR &  [5][6] & OH     &   NIR, MIR, FIR & [6][7]\\
CN     &  sub-mm, mm  & [8] [9] & $^{13}$CN & mm & [53] \\
C$^{15}$N     & sub-mm & [10] & CO     &   NIR, MIR, FIR, sub-mm, mm & [8] [11] \\
$^{13}$CO &  NIR, FIR, sub-mm, mm & [8] [12] & C$^{17}$O &  NIR, sub-mm & [9] [13] \\
 C$^{18}$O &  NIR, sub-mm, mm & [12] [14] & $^{13}$C$^{18}$O &  sub-mm & [14] \\ 
 $^{13}$C$^{17}$O & sub-mm  & [15]  & NO & submm & [54] \\
 CS        &   mm & [8] [17] & $^{13}$CS & mm &  [17] [18] \\
C$^{34}$S &  mm  & [8] [17] & SO        &  sub-mm, mm & [19] [20]\\
\hline
H$_2$O  &   NIR, MIR, FIR, sub-mm &  [7] [21] [22] &  HDO & mm & [55] \\
H$_2^{18}$O & mm & [55] & C$_2$H     & mm & [8] [24]\\
C$_2$D     &  mm & [18] & HCN     &   NIR, MIR, sub-mm, mm & [23] [24]  \\
H$^{13}$CN  & sub-mm, mm &  [25] [23] & HC$^{15}$N     & sub-mm &  [23] \\
DCN         &  mm  & [26] [27] & HNC         &   mm & [8] [28]\\
DNC         &   sub-mm & [18] & HCO$^+$   & sub-mm, mm & [8] [29] \\
DCO$^+$  &   sub-mm, mm & [30] [27] & H$^{13}$CO$^+$   &  sub-mm, mm & [30] [25]\\
HC$^{18}$O$^+$ & sub-mm & [31] & N$_2$H$^+$  & mm & [32] [33]  \\
N$_2$D$^+$   &  mm & [34] [35] & H$_2$S    & mm  & [36] [37]      \\
CO$_2$    &  MIR  & [16][56] &  $^{13}$CO$_2$ & MIR & [56]\\
C$_2$S & mm &  [53] & OCS & mm & [60] \\
SO$_2$   & sub-mm  & [38] & $^{34}$SO$_2$  & sub-mm &  [38]  \\
\hline
NH$_3$  &  MIR, sub-mm & [39][58] & C$_2$H$_2$  &  NIR, MIR & [40] [56] \\
$^{13}$C$^{12}$CH$_2$ & MIR & [57] & H$_2$CO  & sub-mm, mm & [8] [41] \\
H$_2$CS  & sub-mm, mm & [17] [18] & HC$_3$N   & mm  & [42] [43]\\
\hline
HCOOH  &  mm & [44] & $c-$C$_3$H$_2$ & mm & [45] [43]\\
CH$_4$  & NIR & [46] & CH$_2$CN & mm & [47] \\
\hline
CH$_3$CN  & mm & [48] [43] & CH$_3$OH  & sub-mm, mm & [49] [50] \\
C$_4$H$_2$ & MIR & [57] & CH$_3$CHO & submm & [52]\\
CH$_3$OCHO & submm & [51] [52] & $^{13}$CH$_3$OCHO & mm & [60]\\ 
CH$_3$O$^{13}$CHO & mm & [60] & CH$_3$OCH$_3$ & submm & [51] [59]\\
CH$_3$COCH$_3$ & submm & [52] & $t-$C$_2$H$_3$CHO & mm & [60] \\ $s-$C$_2$H$_5$CHO & mm & [60] & C$_6$H$_6$ & MIR & [57]\\
\hline
 \end{tabular}
 \tablenotetext{a}{Adopted and revised from \citet{McGuire2018}. This table includes molecules detected in the disk of V883 Ori, which is a Class I object in outburst phase (section 4.2). The range of wavelengths are 2-5 $\mu$m for near-indrared (NIR), 5-40 $\mu$m for mid-infrared (MIR), 40-300 $\mu$m for far-infrared (FIR), 300-1000 $\mu$m for sub-millimeter (sub-mm), and $\ge 1$ mm for millimeter (mm). References are not exhaustive. In addition to the first detection papers, we list recent papers for readers to refer to references therein (see also text).  \\
References: [1] \citet{weintraub00} [2] \citet{Martin-Zaidi2007} [3] \citet{Bergin2013} [4]\citet{McClure2016} [5] \citet{Thi2011} [6] \citet{Fedele2013} [7] \citet{Salyk08} [8] \citet{Dutrey1997} [9] \citet{Guilloteau2013}  [10] \citet{Hily-Blant2017} [11] \citet{Najita2003} [12] \citet{Ansdell2016} [13] \citet{Qi2011} [14] \citet{Zhang2017} [15] \citet{Booth2019} [16] \citet{Carr08} [17] \citet{LeGal2019} [18] \citet{Loomis2020} [19] \citet{Fuente2010} [20] \citet{Booth2018} [21] \citet{Carr2004} [22] \citet{Hogerheijde2011} [23] \citet{Hily-Blant2019} [24] 
 \citet{Guzman21_MAPS} [25] \citet{Huang17}   [26]  \citet{Qi2008} [27] \citet{Oberg21_TWHya}  [28] \citet{Long2021} [29] \citet{ Aikawa21_MAPS} [30] \citet{vanDishoeck2003} [31] \citet{Furuya22b} [32] \citet{Qi2003} [33] \citet{qi2019} [34] \citet{Huang2015}  [35] \citet{cataldi21} [36] \citet{Phuong2018}  [37] \citet{Riviere2021}  [38] \citet{Booth2021b} [39] \citet{Salinas2016} [40] \citet{Lahuis2006} [41] \citet{Pegues2020} [42] \citet{Chapillon2012} [43] \citet{Ilee21} [44] \citet{Favre2018} [45] \citet{Qi_C2H_2013} [46] \citet{Gibb2013} [47] \citet{canta21} [48] \citet{oberg2015} [49] \citet{Walsh2016} [50] \citet{vanderMarel2021} [51] \citet{Brunken2022} [52] \citet{Lee2019}
 [53] \citet{Phuong21} [54] \citet{leemker23} [55] \citet{Tobin23} [56] \citet{Grant23}
[57] \citet{Tabone23} [58] \citet{Najita21} [59] \citet{Yamato24b}
[60] \citet{Yamato24a}
 }
 \end{small}
\end{table*}

\subsection{Molecules in cold gas}
Fig. \ref{fig:radial_dist} shows the radial distributions of the column densities of gaseous molecules and dust grains towards 3 disks around T Tauri stars (IM Lup, GM Aur, and AS 209) derived by ALMA observations \citep{oberg21} with spatial resolutions of $\sim 18-48$ au.
In the following, we overview what these data tell us about chemistry in disks. Besides the overall chemical structure described above (i.e. the vertically layered structure and snowlines), coupling and decoupling of gas and dust are of special importance, since they could modify not only the molecular abundances but also the local elemental abundances.

\subsubsection{CO and H$_2$O}
In molecular clouds, CO and H$_2$O ice are major Carbon and Oxygen carriers with an abundance of $\sim 10^{-4}$ relative to H$_2$, and thus are naively expected so in the disks. While H$_2$O snowlines are too close ($<$ several au) to the central star to be spatially resolved in the observations, CO snowline is at $r\sim 10-30$ au around T Tauri stars (\S 2.2).
The observation of CO snowline is, however, not straightforward, since CO gas is abundant in the warm molecular layer even outside the CO snowline. N$_2$H$^+$ line is used to trace the CO snowline; its distribution anti-correlates with CO, since it is destroyed (N$_2$H$^+$ + CO $\rightarrow$ N$_2$ + HCO$^+$) when CO is abundant in the gas phase \citep{qi2013,qi2019,Aikawa2015}.
Alternative tracers are rare isotopes of CO (e.g. $^{13}$C$^{18}$O), whose emission is expected to be optically thin \citep{Zhang2017}. In TW Hya, for example, the CO snowline is estimated to be $28-31$ au and 20.5 au from N$_2$H$^+$ and $^{13}$C$^{18}$O, respectively \citep{qi2013,Zhang2017}.

The observations of rare isotopes of CO also suggest that the gaseous abundance of CO is actually lower than $10^{-4}$ even in the warm molecular layer. The CO column densities derived from the observations are compared with the disk models which are constrained by dust continuum emission. As an example, Fig. \ref{fig:radial_dist} shows the radial distribution of CO and dust column densities derived from C$^{18}$O ($J=2-1$) and dust continuum \citep{zhang21,Sierra21}.
Assuming the gas/dust mass ratio is 100, the CO depletion factor, i.e. the ratio of canonical CO abundance ($10^{-4}$) divided by the estimated CO abundance in the warm molecular layer, varies from 1 to 100 among disks in radii outside the CO snowline \citep{Miotello2017,Zhang2019,zhang21}. While the CO abundance tends to increase inwards, the CO depletion factor is $10-100$ even inside the CO snowline in some disks.
An alternative interpretation of these observations is that the gas/dust mass ratio is lower than 100. The submillimeter HD emission lines, which are observed by {\it Herschel} towards a limited number of disks, suggest CO depletion rather than gas depletion \citep{Bergin2013,Favre2013,McClure2016}. It should however be noted that the evaluation of disk gas mass from HD emission also depends on the assumed thermal structure, i.e. disk models \citep{Miotello23}. 

{\it Herschel} also observed a few H$_2$O lines in several disks
\citep{Hogerheijde2011,Salinas2016, Du2017}. While the beam size (e.g. 37$"$ at 557 GHz) is too large to provide any information on the spatial distribution, the emission lines of H$_2$O are significantly fainter than the predictions of disk models, in which H$_2$O is photodesorbed from ice outside its snowline. It suggests that H$_2$O ice-coated grains are mostly sedimented to the disk midplane where photodesorption is inefficient, and could even be radially drifted inwards (\S 2.3). 
The low CO abundance in the warm molecular layer can be explained similarly by the cold-finger effect; while CO is in the gas phase in the warm layer, it could be brought to the cold midplane by weak turbulence to freeze out onto grains. Conversion of CO to less volatile molecules (e.g. CO$_2$ ice and CH$_3$OH ice) would also help \citep[e.g.][]{Furuya2014, Kama2016,Krijt2020, Furuya22a}.

It is noteworthy that H$_2$O ice and CO are expected to be the dominant reservoirs of oxygen and carbon in the initial condition (i.e. gas and solids accreted from the molecular clouds). While both can be depleted from the warm molecular layer as described above, lower volatility of H$_2$O would result in heavier depletion of oxygen than carbon, and an elemental abundance of C/O $>1$ in the warm molecular layer. The high C/O ratio is supported by the observations of other molecules as explained below.

The similarity of the cometary ice with interstellar ice may indicate the sedimentation of ice-coated grains before ice composition is significantly affected by UV and X-rays in the upper layers of the Solar nebula, while molecules formed in the upper layers can still be incorporated into the midplane ice via the cold finger effect.

\subsubsection{Hydrocarbons: C$_2$H and $c$-C$_3$H$_2$}
Unsaturated hydrocarbons, C$_2$H and $c$-C$_3$H$_2$, are expected to be abundant in the surface PDR layer, which is supported by the imaging observation of edge-on disks \citep{Ruiz-Rodrigurz_2021} and their relatively high excitation temperature (20 K $-$ 50 K)
\citep{Guzman21_MAPS, Cleeves2021}. 

The abundance of C$_2$H is also sensitive to the C/O ratio in the gas phase; the comparison of its line brightness between observations and models indicates C/O $>1$ \citep{Bergin2016,Bergner2019,Miotello2019}. However, variation among disks should be noted; some disks are faint in C$_2$H emission \citep{Miotello2019}. Spatially resolved observations show that the C$_2$H emissions often show a ring-like structure. Some rings coincide with the prominent gaps seen in the dust continuum, which is consistent with the expectation that C$_2$H abundance is enhanced by UV penetration. But not all dust gap correlates with C$_2$H ring, and vice visa \citep{Guzman21_MAPS,Law21_MAPS_radial_profiles}.

$c$-C$_3$H$_2$ is detected in several disks, although its emission lines are weaker than those of C$_2$H \citep{Qi_C2H_2013}. The column density ratio of $c$-C$_3$H$_2$/C$_2$H is relatively constant with $5-10 \%$ in four disks where this ratio is derived \citep{Ilee21} (Fig. \ref{fig:radial_dist}). These observations and theoretical models suggest that $c$-C$_3$H$_2$ is formed in the gas phase together with C$_2$H.

\subsubsection{HCN and CN}
CN is also predicted to be abundant in the surface PDR layer, which is supported by the excitation analysis \citep{Teague_2020, Bergner21_MAPS}. A ring-like distribution of CN emission is found in some disks, which is explained by CN formation via the reaction of N atom with UV-pumped vibrationally excited H$_2$ \citep{Cazzoletti2018_CN}. HCN, on the other hand, is more susceptible to photodissociation than CN; it is dissociated by UV at longer wavelengths than for CN, and by Ly$\alpha$ emitted from T Tauri stars. 

A recent high-resolution observation by \citet{Bergner21_MAPS} show that the column densities of CN and HCN are positively correlated (Fig. \ref{fig:radial_dist}), which suggests that their formations are connected. For example, chemical models predict that both CN and HCN abundances are enhanced if the C/O ratio is $>1$ in the gas phase \citep{Cleeves2018, Cazzoletti2018_CN}. The column density ratio of CN/HCN also increases with radius (Fig. \ref{fig:radial_dist}), which is consistent with the model prediction that CN is more abundant in the lower density UV-irradiated disk surface \citep{Bergner21_MAPS}.

\subsubsection{S-bearing molecules}
As in molecular clouds, the abundances of gaseous S-bearing molecules in the disks are much lower than expected from model predictions assuming the Solar elemental abundance, which suggests significant depletion of sulfur from the gas to the solid phase (see \S 4.2 and the chapter by Bergin et al.). Six S-bearing molecules are detected in disks: CS, SO, H$_2$S, H$_2$CS, C$_2$S and SO$_2$. Among them, CS is the most readily detected, and its column density is estimated to be $10^{12}-10^{13}$ cm$^{-2}$ in several disks \citep{LeGal2019}. Detection of other S-bearing molecules is rather limited.
SO is first detected in AB Aur disk by \citet{Fuente2010}. \citet{Guilloteau2016} observed SO in 20 disks using IRAM 30 m and detected the emission only in 4 disks. Prevalence of CS over SO may suggest a high C/O ratio in disks \citep{LeGal21}. H$_2$S, the dominant sulfur-bearing molecule in comets, is first detected in disks by \citet{Phuong2018}, and is recently detected in 4 additional disks \citep{Riviere2021}. H$_2$CS is detected in the disk of Herbig Ae star, MWC 480 \citep{LeGal2019}. 

Recently SO$_2$ was first detected in the disk of IRS 48, which is a transitional disk (i.e. a disk with a central hole in dust continuum emission) with an asymmetric dust emission peak  \citep{Booth2021b}. The emission peak would correspond to the local pressure maximum, at which dust grains are trapped (\S 2). Both SO and SO$_2$ are bright only at this dust peak, while CS is not detected in this disk. The low CS/SO ratio indicates a low C/O ratio ($<1$) at the dust trap of IRS 48.

\subsubsection{Complex Organic Molecules}
In the astrochemistry community, organic molecules consist of 6 atoms and more are called complex organic molecules (COMs). Methanol, the most abundant and prototypical COM in molecular clouds, is so far detected only in TW Hya, HD 100546, IRS 48, and HD 169142 \citep{Walsh2016,Booth2021a, vanderMarel2021, Booth2023} in spite of deep searches in some other disks \citep{Loomis2020,Carney2019}. It is also the key molecule to investigate the relation between ice and gas in disks; since its formation in the gas phase is known to be inefficient, gaseous CH$_3$OH should be mainly desorbed from ice, which in turn, is formed by hydrogenation of CO on cold ($\le 20$ K) ice surface. The CH$_3$OH emission in TW Hya is very weak and is in a ring region outside the CO snowline ($\sim 30$ au) and inside the millimeter dust continuum edge. Since the dust temperature in the emitting region is below its sublimation temperature ($\sim 100$ K), CH$_3$OH is considered to be desorbed by non-thermal processes. HD 100546, on the other hand, is a warm Herbig Ae disk with a central hole in the dust continuum. Most of the CH$_3$OH emission originates from the spatially unresolved central region ($\le 60$ au) and thus could be tracing the thermally sublimated CH$_3$OH. The warm dust temperature ($>20$ K) in the disk of HD 100546 indicates that CH$_3$OH is not formed in the disk, but inherited from molecular clouds \citep{Booth2021a}.
In IRS 48, the CH$_3$OH emission spatially coincides with the dust trap and shows a high excitation temperature $\sim 100$ K. Gaseous methanol abundance could be enhanced by a combination of ice-coated pebble concentration in the dust trap and irradiation heating at the edge of the central hole \citep{vanderMarel2021}. Recently, CH$_3$OCH$_3$ and CH$_3$OCHO are also detected at the same position as CH$_3$OH in IRS 48 \citep{Brunken2022}.

\begin{figure*}[ht!]
\centering
\resizebox{0.8\hsize}{!}{\includegraphics{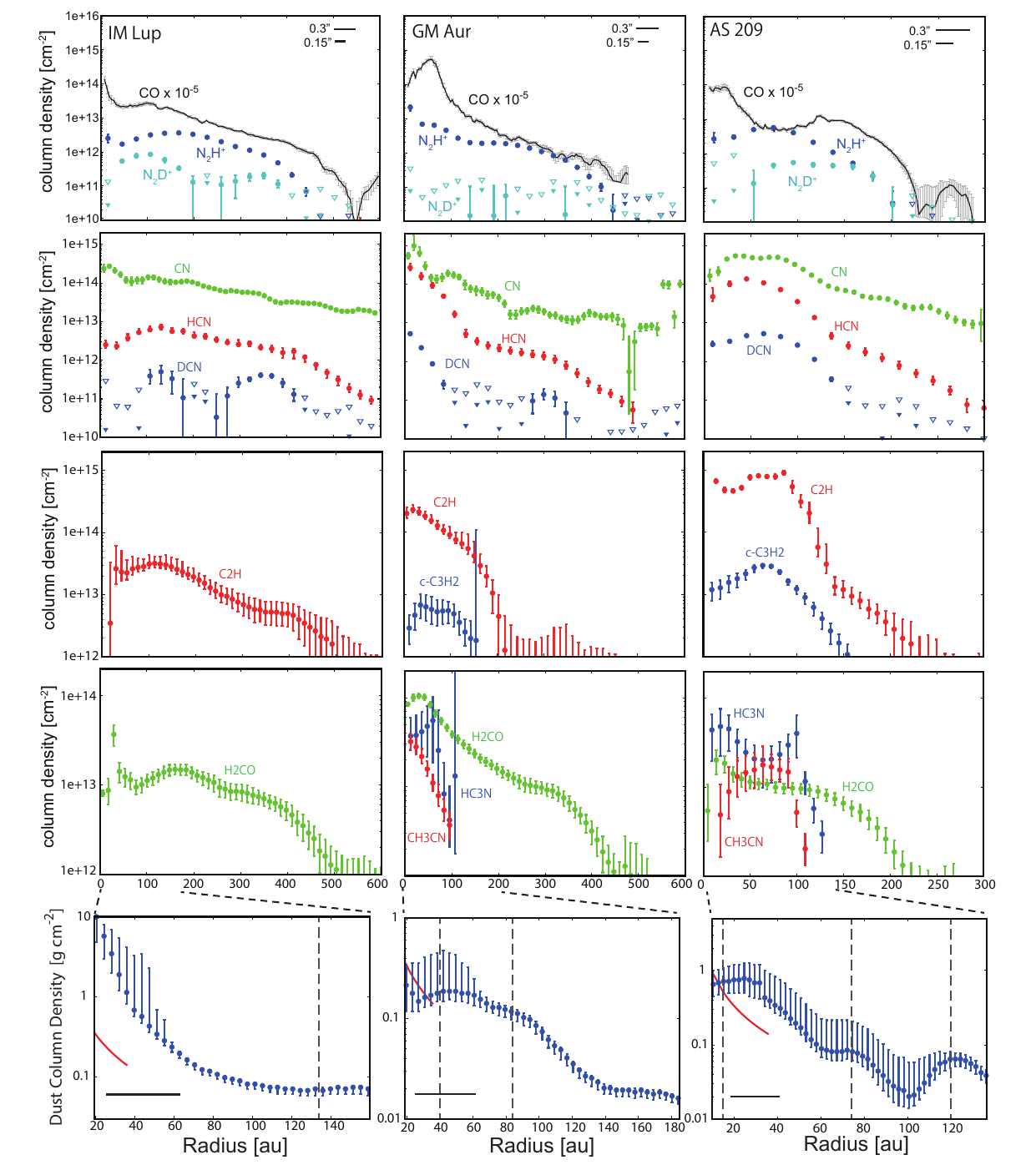}}
\caption{Radial column density distributions of gaseous molecules and dust column densities in IM Lup (left), GM Aur (middle), and AS 209 (right) from \citet{zhang21, Bergner21_MAPS,cataldi21,Guzman21_MAPS,Ilee21} and \citet{Sierra21}. The molecular column densities of C$_2$H and CO are derived from the data with a beam size of $0.15"$, while the beam size is $0.3"$ for other molecules. The error bars correspond to 1 sigma. For N$_2$H$^+$, N$_2$D$^+$, and DCN, the 1-sigma and 3-sigma upper limits are plotted at radii where the median value of the molecular column density is lower than the 1-sigma upper limits by a factor of $> 10$. In the panels of dust column densities, the resolution of the dust continuum observations is shown with a horizontal bar, while the vertical dashed lines depict the positions of the rings. The red curves depict the dust column density of the MMSN (see Section~\ref{sec:substructures}).
\label{fig:radial_dist}.}
\end{figure*}

Formic acid (HCOOH), which is often detected in the protostellar cores with bright CH$_3$OH emissions, is so far detected only in TW Hya \citep{Favre2018}.
H$_2$CO is another relevant molecule to CH$_3$OH. It is an intermediate product of CO hydrogenation to form CH$_3$OH on grain surfaces, while it can be formed via gas-phase reactions as well. H$_2$CO is also considered to be a precursor species of organics in meteorites \citep{cody11}.
\citet{Pegues2020} detected H$_2$CO in 13 disks out of 15, and derived an excitation temperature of $20-50$ K and column density of $\sim 5\times 10^{11}-5\times 10^{14}$ cm$^{-2}$. The second bottom panels in Fig. \ref{fig:radial_dist} show the radial distribution of H$_2$CO in 3 T Tauri disks assuming the excitation temperature of 20 K \citep{Guzman21_MAPS}.

A nitrogen bearing COM, CH$_3$CN, and another nitrile species HC$_3$N are detected in several disks \citep{oberg2015,Bergner2018}. \citet{Ilee21} spatially resolved the emission lines of these nitriles in disks of GM Aur, AS 209, HD 163296, and MWC 480; the emission is compact ($\le 100$ au) comparable to the extent of the dust continuum. Unlike CH$_3$OH, these species can be formed in the gas phase. The weak correlation of their emission distributions with the dust continuum, however, may indicate the role of ice reservoir for their formation. The excitation temperatures ($\sim 30-50$ K) are lower than their sublimation temperatures, indicating that they are non-thermally desorbed from ice or formed in the gas phase from the species with lower sublimation temperatures. Their column densities ($10^{13}-10^{14}$ cm$^{-2}$) (Fig. \ref{fig:radial_dist}) are higher than predicted in static disk models; either or a combination of grain growth and turbulent mixing would enhance their abundances \citep{Semenov11,Furuya2014,oberg2015}.

\subsection{Molecules in warm and hot gas}
Abundant warm molecular gas ($T\gtrsim 200\,$K) is generally observed in typical protoplanetary disks surrounding stars of widely different masses \citep{Brittain03, Blake04, Brown13}. The regions of protoplanetary disks where such warm gas is present are often referred to as the ``inner disk''. The inner disk may similarly be defined as corresponding to disk radii inside the water snowline. While refractory carbon and silicate-dominated dust sublimate at 1500\,K, i.e. at the radius of 0.01-0.1\,au for typical protoplanetary disks, many chemically important volatile molecular species (water, CO$_2$, CH$_3$OH, CH$_4$, NH$_3$, etc.) persist in the gas phase down to dust temperatures of $\sim$100\,K or below. Such temperatures are achieved at 1-2\,au in the midplane for disks around young solar-mass stars, and at distances of up to $\sim$10\,au in the superheated disk surface \citep{Sasselov00,Garaud07} (Fig, \ref{fig:Tmid}). Indeed, high-resolution spectroscopy and spectral imaging have demonstrated that the emission originates well inside their respective snowlines \citep{Goto06,Pontoppidan08}, and that the location of the snowline likely varies with the evolutionary stage of the star-disk system \citep[][see also \S 2.2]{Hsieh19}. 

\begin{figure*}[ht!]
\centering
\includegraphics[width=13cm]{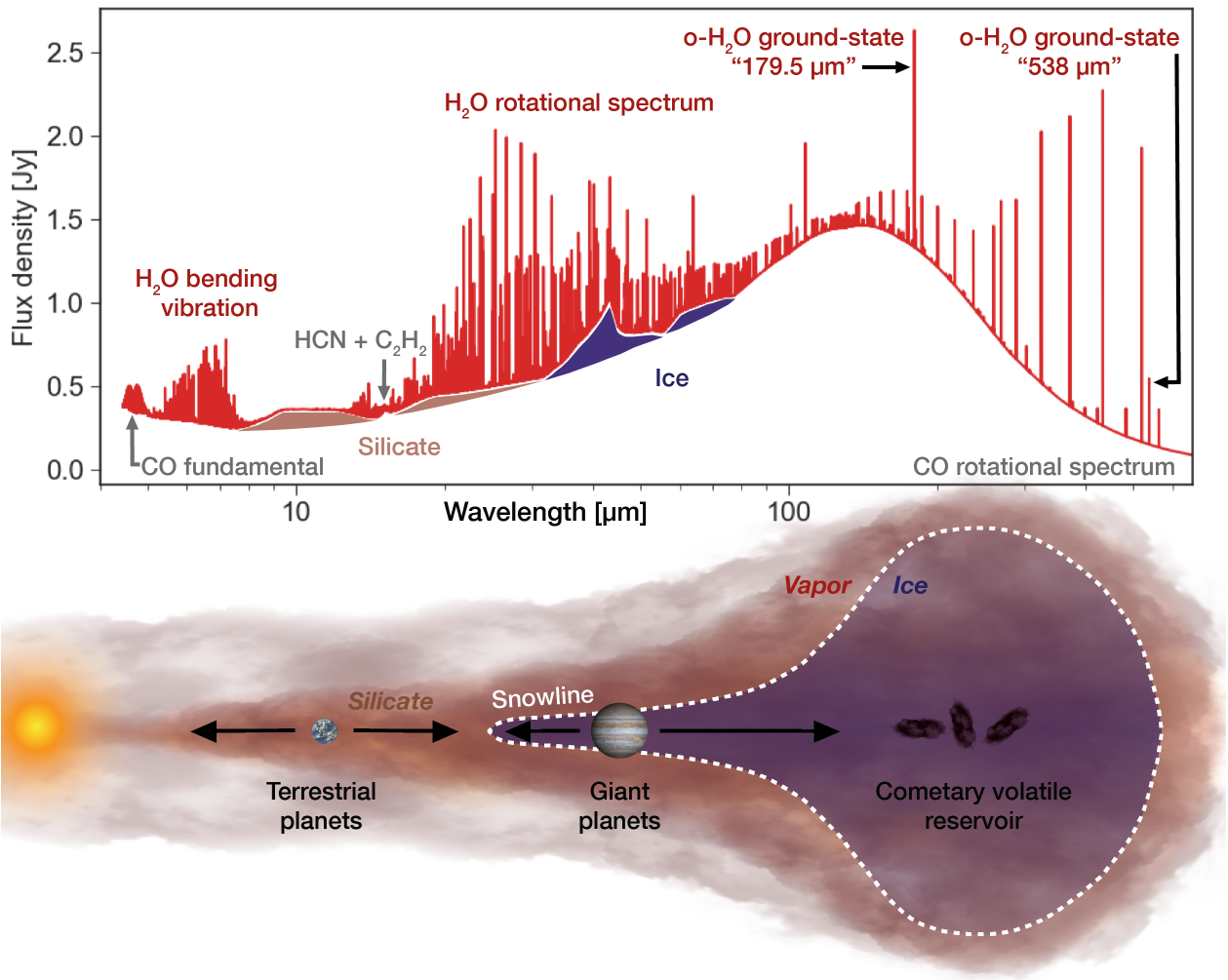}
\caption{Radiative transfer model infrared spectrum of a water-rich protoplanetary disk around a solar-mass star, based on observed {\it Spitzer} and {\it Herschel} spectra. The spectrum is rendered at high spectral resolving power ($\rm 3\,km\,s^{-1}$). The spectrum includes lines from CO, water, and organics, and also includes the far-infrared solid-state bands due to water ice at 43 and 62\,$\mu$m. The figure is reproduced from \cite{Pontoppidan19a}.}
\label{fig:ir_disk_spectrum}
\end{figure*}

Warm molecular gas in disks is readily observed at mid-infrared wavelengths through their ro-vibrational transitions, through high-J pure rotational transitions in the case of water, and in the ultra-violet through electronic transitions (Fig. \ref{fig:ir_disk_spectrum}). It is also possible to observe warm and hot gas through rotational transitions between excited vibrational states in the (sub)millimeter regime, but beam dilution of the typically very small emitting inner disk regions makes such lines challenging to detect and image, even with facilities such as ALMA \citep{Notsu19}. 

Given the theoretical prediction that the water snowline plays a key role in the formation of planets, multiple attempts have been made to observationally measure its location (and that of similar species), but with mixed results. While the CO snowline is located at disk radii large enough to allow for measurements using spatially resolved millimeter interferometric imaging (\S 3.1.1), the water snowline typically only subtends an angle of 10-100 milli-arcseconds, below the imaging resolution limit of current facilities. Further, while some hot water lines are detectable from the ground, water can only reliably and consistently be detected from space. Consequently, various techniques using purely spectral signatures have been used to constrain the location of the water snowline. \citet{Zhang13} and \citet{Blevins16} used the line spectral energy distribution, representing gas at a range of temperatures, to constrain
the water sublimation radius in the surface layer
in five protoplanetary disks, finding values of 3-10\,au. \cite{Notsu16} proposed a method using optically thin water lines from
transitions with low spontaneous emission probability
to probe the snowline deeper in the disk, but attempts at detecting the intrinsically weak lines have so far not been successful \citep{Notsu19}. 

Because typical protoplanetary disks are primarily externally heated, except for the innermost regions ($<1$\,au) or disks with very high accretion rates ($>10^{-7}\,M_{\odot}\,yr^{-1}$) (\S 2.2), warm molecular gas lines in the infrared overwhelmingly trace the surface layers of the disks \citep{Kamp04}. This effect is exacerbated by the high vertical optical depths of most disks in the 1-10\,au radius range. Consequently, the composition of the disk midplanes at these radii is generally not being observed, and can only be inferred under the assumption that some degree of vertical mixing is present \citep{Semenov11,Anderson21}. 

The relative abundances of warm gas-phase volatiles in inner disk surfaces appear to be different from those observed in ices in the cold interstellar medium, and in Solar system comets. This indicates that the observed inner disk chemistry is not directly inherited from cold ice chemistry, but is significantly altered by local processes. Concurrent measurements of H$_2$ and CO using UV spectroscopy suggest that inner disk CO abundances range from slightly depleted up to canonical ([CO/H$_2$]$\sim$10$^{-5}$--10$^{-4}$) \citep[][]{France14,Cauley21}. Other observable species, such as NH$_3$, CH$_4$, and CO$_2$, are further depleted relative to CO by orders of magnitude, clearly inconsistent with primordial ice chemistry \citep{Mandell12, Pontoppidan14, Bosman17, Pontoppidan19}, but consistent with current predictions for warm gas-phase chemistry, which tends to destroy particularly those species, driving nitrogen into N$_2$ and carbon into CO \citep{Agundez08,Walsh15}. Water, on the other hand, is commonly highly abundant in disks around low-mass and solar-mass young stars, as revealed by an extensive sample of disks observed by {\it Spitzer} \citep{Carr08,Salyk08,Pontoppidan2010,Carr11}, with retrieved abundances consistent with [CO/H$_2$O]$\sim$1 \citep{Salyk11}. 
JWST is currently confirming the ubiquitous presence of mid-infrared emission from warm gas in inner disks around young stars of all masses, and demonstrate a widening diversity of relative molecular abundances from water and organics \citep{Tabone23,Banzatti23,vandishoeck23}.

\subsection{Observations of ices in disks}

During most of the evolution of a planetary system, the bulk of its volatile component is sequestered in the solid phase in the form of molecular ices. Indeed, most of the disk mass is typically located beyond the snowline (Fig. \ref{fig:ir_disk_spectrum}), with a few notable exceptions such as disks undergoing violent outbursts from instability-induced accretion events, or disks that have been truncated by stellar companions. 

While gas-phase volatiles can be relatively easily observed in the warm disk surface, or the hot inner disk inside the snowline, it has proven to be challenging to observe ices in disks, especially in their cold, outer midplanes where the formation of comets is presumably active. Ices are most readily identified through their strong mid-infrared (3-20\,$\mu$m) resonance bands. However, these bands are nearly universally seen in absorption toward a background source of light, as the dust temperatures required to excite the bands in emission are too high ($>150\,$K) to retain the ice on the emitting grains. Further, the midplanes of protoplanetary disks during planetesimal formation are generally hidden beneath highly optically thick (at infrared wavelengths) layers of dust. 

As a result, direct observations of ices in protoplanetary disks are sparse, and rely on particular geometric configurations that tend to be difficult to interpret. Even in cases where disk ices are detected, such ice is confined to the surface layers of the disk at $\tau_{\rm IR}\sim 1$, corresponding to a fraction of a \% of the total vertical column density. 

Ices in disks have been observed when the disk is viewed close to edge-on, where ice in a flared outer disk absorbs thermal emission from hot dust in the inner disk, light scattered off the disk surface, or some combination thereof \citep{Pontoppidan05,Terada07}. Interpretation of such direct absorption spectroscopy is complicated by the complex geometry of the radiative transfer, making it difficult to establish unambiguously the location of the ices within the disk. Further, the signal-to-noise ratios of edge-on disk spectra have often been low due to the inherent faintness of edge-on disks at most infrared wavelengths. In many cases, part of the absorption may also be due to cold material in a remnant protostellar envelope, rather than in the disk itself. In the most unambiguous cases, it appears that the most abundant ices have relative concentrations consistent with those observed in dense clouds and protostellar envelopes \citep{Aikawa12}, although more sensitive observations of rarer species, such as CO$_2$ and CH$_3$OH, will likely be needed to establish evolutionary patterns \citep{Ballering21}. 

A complementary method for detecting the mid-infrared resonances of ice is through their detection in scattered light. This method is limited for use at the shortest wavelengths where scattering is the most efficient for the grain sizes present in disk surfaces ($<5 \, \mu$m). Thus far, detections of the 3\,$\mu$m water ice stretching mode have been made in medium-band coronagraphic imaging of a number of face-on disks \citep{Inoue08, Honda09,Honda16}. While clearly demonstrating the presence of ice-coated grains in the uppermost disk surface layers, model fits to ice band optical depths were not able to accurately measure the surface ice/rock ratio, and in particular whether photodesorption is playing a significant role in depleting ice in photo-dominated regions of the disks. 

As opposed to the strong mid-infrared ice resonances at 3--15\,$\mu$m, the far-infrared lattice vibration modes of particularly water ice near 43 and 62\,$\mu$m \citep{Smith84,Hudgins93} appear {\it in emission} from a significant fraction of the disk dust mass in the comet-forming regions \citep{Malfait98}. The excitation of these features in low-temperature dust means that they are potentially powerful tracers of the ice mass reservoirs. Further, as they are related to vibrations of the solid lattice structure, rather than to vibrations within individual molecules, they are highly sensitive tracers of ice phase and crystallinity. The far-infrared water ice bands have been detected in at least 4 protoplanetary disks, and suggest high abundances of water ice, yielding estimates of ice/rock mass ratios of 0.36-1.6 \citep{McClure15,Min16}. Attempts have been made to detect far-infrared features in disks from ice species other than water, but these have thus far been unsuccessful, likely due to limitations of sensitivity \citep{McGuire16,Giuliano16}. 

Observations of the molecular inventory in the inner disk may be used to constrain the efficiency of radial transport of ices. \cite{Najita13} found a trend between the mid-infrared HCN/H$_2$O flux ratio and dust disk mass as measured by ALMA. The proposed interpretation of this apparent link between inner and outer disk properties is that more massive disks may have experienced more efficient build-up of large, water-rich planetesimals, leaving less water ice available in dust grains that migrate inwards. In this case, the observed HCN/H$_2$O ratio is also a measure of the inner disk elemental C/O ratio. More recently, this scenario was supported by the observation of \cite{Banzatti20}, who found a similar relation of enhanced water emission for disks with small dust emission radii, suggesting that compact disks have experienced strong inward radial drift of water-ice rich dust, thereby enriching the inner disk with water vapor and decreasing the local C/O ratio (see also \S 4.2). 

\subsection{key markers for comets}

\subsubsection{Isotope Fractionation}

In the Solar system material, isotope ratios of molecules are often different from the elemental abundance ratio. It is called isotope fractionation, and is used to investigate the origin of the planetary material. Isotope ratios in comets and other Solar system materials are reviewed in the chapter by \citet{Biver2023}. Here we summarize the observations and models of fractionation in disks. We refer \cite{nomura23} for a more detailed review. There are two mechanisms known to be responsible for the fractionation: (i) exothermic exchange reactions at low temperatures and (ii) selective photodissociation.

The D/H ratio of Earth's ocean ($1.56\times 10^{-4}$) is higher than the elemental abundance of D/H in the interstellar matter ($1.5\times 10^{-5}$) or the protosolar value ($2.1\times 10^{-5})$ (see the chapter by Biver et al. and references therein). Such deuterium fractionation (enrichment) is caused by exchange reactions, e.g.
\begin{eqnarray}
{\rm H_3^+ + HD \rightarrow H_2D^+ + H_2} \label{H2Dp} \\
{\rm CH_3^+ + HD \rightarrow CH_2D^+ + H_2.} \label{CH2Dp}
\end{eqnarray}
Since these reactions are exothermic, the backward reactions are inefficient at low temperatures ($\le$ several 10 K). The high D/H ratio (e.g. H$_2$D$^+$/H$_3^+$) propagates to other molecules via chemical reactions; e.g. HDO is formed by grain surface reactions of O atom or OH with D atom, whose abundance is enhanced by the dissociative recombination of H$_2$D$^+$.
The high D/H ratio of the ocean suggests that water is supplied to Earth from cold regions, at least partially, and has been motivating the measurement of D/H ratios in comets. The D/H ratio of cometary water, $(1.4-6.5)\times 10^{-4}$, overlaps with that of ocean \citep[][Chapter by Biver et al.]{Mumma2011,ceccarelli14,Altwegg2015}.
Recent observations of higher D$_2$O/HDO ratio than HDO/H$_2$O both in comet 67P/C-G and protostellar cores indicate the inheritance of comet water from molecular clouds \citep{Furuya2016,Furuya2017,Altwegg2017}, while the variation of HDO/H$_2$O ratio among comets indicates some reprocessing in disks.

In protoplanetary disks, five deuterated molecules, DCO$^+$, N$_2$D$^+$, DCN, C$_2$D, and HDO are detected so far \citep[e.g.][]{Huang17, Salinas17,cataldi21,Oberg21_TWHya,Loomis2020,Tobin23} (Table \ref{tab:molecules}).
Disk averaged D/H ratio of the former three species are summarized in Table \ref{tab:DtoH}.
It should be noted that the recombination timescale of ions is much shorter than the typical ages of disks (i.e. a few $10^6$ yr). The high D/H ratio of molecular ions is thus clear evidence of active deuterium fractionation in disks.
The D/H ratio of N$_2$H$^+$ is higher than that of HCO$^+$. It is consistent with the theoretical expectation that N$_2$H$^+$ traces the cold ($<20$ K) layer near the midplane where CO is frozen out, which enhances the deuterium fractionation \citep{Willacy2007,Cleeves2014, Aikawa2015,Aikawa2018}. 

Recently, \citet{Tobin23} detected HDO and H$_2^{18}$O in the disk of FU Ori star V883 ori, and derived the HDO/H$_2$O ratio to be $(2.23\pm 0.63) \times 10^{-3}$. Since this ratio is comparable to those observed in the warm central protostellar cores, they conclude that  water in the disk inherit from molecular clouds.

Since temperature increases radially inwards in disks, the molecular D/H ratio is expected to decrease inwards, which is confirmed by the spatially resolved observations of DCN, DCO$^+$ and N$_2$D$^+$ \citep{cataldi21,Oberg21_TWHya}. The observed peak D/H ratios and their radial positions are summarized in Table \ref{tab:DtoH} (see also Figure \ref{fig:radial_dist}).
The radial distribution of the molecular D/H ratio could also probe the deuteration pathways of each molecular species. The decline of D/H ratio towards the warm central region is expected to be less steep if the molecule is formed mainly via hydrocarbons rather than H$_3^+$, since the endothermicity of reaction (\ref{CH2Dp}) is higher than that of (\ref{H2Dp}) \citep{Oberg2012}.
\citet{cataldi21} derived the DCN/HCN ratio as a function of the excitation temperature of HCN (i.e. proxy of gas temperature), which indicates the significant contribution of reaction (\ref{CH2Dp}). They also found that the DCN/HCN ratio becomes comparable to the value measured in Hale-Bopp, $(2.3\pm 0.4)\times 10^{-3}$ \citep{Meier1998,Crovisier2004}, in the regions of $\sim 30-40$ K.

The $^{14}$N/$^{15}$N ratio of HCN, CN, NH$_2$ has been measured in comets. The ratio is lower than the elemental abundance in the local interstellar medium ($\sim 200-300$) \citep[e.g.][]{Ritchey2015} and the protosolar value (441) \citep{Marty2011}. For example, the average C$^{14}$N/C$^{15}$N ratio over 20 comets is $147.8\pm 5.7$ \citep{Manfroid2009}.
The $^{15}$N enrichment could be due to exchange reactions and/or selective photodissociation of N$_2$. The former would be less effective than previously thought, since some key exchange reactions are found to have activation barriers
\citep{Roueff2015}. In the latter mechanism, $^{14}$N$^{15}$N is photodissociated in deeper layers of a disk or molecular cloud than the major isotope, producing excess $^{15}$N atoms, which are incorporated to other molecules such as CN \citep{Liang2007,Heays2014,Furuya2017}.
\citet{Visser2018} calculated $^{14}$N/$^{15}$N fractionation in disk models to find that the fractionation is fully dominated by selective photodissociation of N$_2$ \citep[see also][]{Lee2021}.

\citet{Guzman2017} observed H$^{13}$CN and HC$^{15}$N to find HCN/HC$^{15}$N ratio of 80-160 in five disks, which are roughly consistent with the values in comets. In the disk of TW Hya (Table \ref{tab:N_isotope}), \citet{Hily-Blant2019} found that the HCN/HC$^{15}$N ratio increases with radius, which is consistent with the fractionation due to selective photodissociation \citep[see also][]{Guzman2017}. The ratio of CN/C$^{15}$N, on the other hand, is $323\pm 30$ \citep{Hily-Blant2019}. The different $^{14}$N/$^{15}$N ratio in HCN and CN could be a natural outcome of the vertical gradient of molecular abundance; i.e. CN has its abundance peak in the upper layer than that of HCN \citep{Cazzoletti2018_CN, Teague_2020} (\S 3.1).

Carbon and oxygen isotope ratios in comets are mostly consistent with the Solar abundance, with the notable exception of some molecules in 67P/C-G (see the chapter by Biver et al. for more details). On the theoretical side, selective photodissociation of CO, combined with isotope exchange reaction of $^{13}$C$^+$ + CO $\rightarrow$ $^{13}$CO + C$^+$, could induce fractionation of $^{12}$C/$^{13}$C and $^{16}$O/$^{17}$O/$^{18}$O \citep{Yurimoto2004, Lyons2005}.
\citet{Miotello2014} calculated the isotope fractionation of CO in disk models to show that the abundance and thus the line flux of C$^{18}$O and C$^{17}$O could be significantly reduced by the selective photodissociation, especially in disks with grain growth. Observational confirmation is challenging, since the lines of the major isotope of CO are optically thick, while the rare isotope lines are very weak.
\citet{Smith2009} found that the C$^{18}$O/CO and C$^{17}$O/CO ratios are lower than the elemental abundances in the disk of VV CrA by observing absorption lines in the near-infrared with the binary star as a possible light source.
As for carbon, 
\citet{Hily-Blant2019} found that the HCN/H$^{13}$CN ratio is mostly flat with $77.9-88.7$ at $R=20-55$ au in the disk of TW Hya. 
\citet{yoshida22}, on the other hand, derived $^{12}$CO/$^{13}$CO$=21\pm 5$ at $70-110$ au, which indicates the fractionation via the isotope fractionation (C$^+$ + CO) in the gas of high C/O ($>1$) elemental abundance.
\cite{Furuya22b} detected HC$^{18}$O$^+$ emission in TW Hya. Combining this detection with previous H$^{13}$CO$^+$ observation and disk chemical model, $^{13}$CO/C$^{18}$O ratio is estimated to be consistent with the elemental abundance ratio in the local ISM.
More recently, \citet{Yamato24a} found that the $^{12}$C/$^{13}$C ratios of freshly sublimated complex organic molecules in V883 Ori is low ($\sim 20-30$) (section 4.2).

In summary, we see at least a qualitative agreement in deuterium enrichment and nitrogen fractionation between disks and comets, in spite of the fact that we are observing gaseous components in disks, rather than ice.

\begin{table*}[h]
\caption{Molecular D/H ratios measured in Class II disks and comets}
\label{tab:DtoH}
 \begin{tabular}{cccc}
 \hline
object & DCO$^+$/HCO$^+$  & N$_2$D$^+$/N$_2$H$^+$ & DCN/HCN   \\
\hline
DM Tau & 0.1 (50 au) $-$ 0.2 (450 au)\tablenotemark{a} &     &           \\
TW Hya & 0.01 (30 au) $-$ 0.1 (70 au)\tablenotemark{b} &     & 0.17\tablenotemark{c}    \\
AS  209 &   0.037-0.08\tablenotemark{d}  & 0.3-0.5\tablenotemark{e}  & 0.028-0.059\tablenotemark{d}  \\
        &              & 0.67-0.94 (159 au)\tablenotemark{f} & 0.074-0.092 (118 au)\tablenotemark{f}  \\
IM Lup  &   0.023-0.037\tablenotemark{d}  &        & 0.043-0.074\tablenotemark{d}   \\
        &               & 0.28-0.37 (76 au)\tablenotemark{f} &   0.17-0.24 (367 au)\tablenotemark{f}  \\
V4046 Sgr & 0.014-0.029\tablenotemark{d} &        & 0.004-0.008\tablenotemark{d}  \\
LkCa15 &  0.019-0.038\tablenotemark{d} &           & 0.04-0.12\tablenotemark{d}  \\
GM Aur &              & $< 0.064$ (216 au)\tablenotemark{f}\ & 0.079-0.13 (298 au) \tablenotemark{f}  \\
MWC 480 & 0.021-0.043\tablenotemark{d} &           & 0.006-0.018\tablenotemark{d}  \\
        &             & 0.40-0.66 (143 au)\tablenotemark{f} & 0.055-0.094 (133 au)\tablenotemark{f} \\
HD 163296 & 0.039-0.075\tablenotemark{d} &         &  0.012-0.027\tablenotemark{d}  \\
          & 0.04-0.07\tablenotemark{g} & 0.19-0.66\tablenotemark{g} & 0.01-0.03\tablenotemark{g} \\
          &            & 1.29-1.72 (152 au)\tablenotemark{f} & 0.048-0.062 (129 au)\tablenotemark{f} \\
\hline
comet Hale-Bopp       &     &   &  $(2.3\pm0.4)\times 10^{-3}$\tablenotemark{h}  \\
\hline
 \end{tabular}
\tablenotetext{a}{\citet{Teague2015}} \tablenotetext{b}{\citet{Qi2008}}
\tablenotetext{c}{disk average \citep{Oberg2012}}
\tablenotetext{d}{disk average \citep{Huang17}}
\tablenotetext{e}{disk average \citep{Huang2015}}
\tablenotetext{f}{peak value in the radial distribution \citep{cataldi21}}
\tablenotetext{g}{disk average \citep{Salinas17}}
\tablenotetext{h}{\citet{Meier1998, Crovisier2004}}
\end{table*}

\begin{table*}[h]
\caption{Molecular $^{14}$N/$^{15}$N ratios measured in Class II disks and comets}
\label{tab:N_isotope}
 \begin{tabular}{ccc}
 \hline
object & HC$^{14}$N/HC$^{15}$N  & C$^{14}$N/C$^{15}$N   \\
\hline
TW Hya & $223\pm 21$\tablenotemark{a} &  $323\pm 30$\tablenotemark{a}    \\
    &    $121\pm 11$ (20 au) $- 339\pm 21$ (45 au)\tablenotemark{b} &  \\
AS  209 &   $156\pm 71$\tablenotemark{c} &    \\
V4046 Sgr & $115\pm 35$\tablenotemark{c} &      \\
LkCa15 &  $83\pm 32$\tablenotemark{c} &       \\
MWC 480 & $123\pm 45$\tablenotemark{c} &        \\
HD 163296 & $142 \pm 59\tablenotemark{c} $ &      \\
\hline
comet Hale-Bopp   & $205 \pm 70$\tablenotemark{d}  &  $140\pm 35$\tablenotemark{e}  \\
comet 17P/Holmes  & $139\pm 26$\tablenotemark{d}  & $165\pm 40$\tablenotemark{d}  \\
\hline
 \end{tabular}
 \tablenotetext{a}{disk average \citep{Hily-Blant2019}}
  \tablenotetext{b}{radial distribution \citep{Hily-Blant2019}}
 \tablenotetext{c}{disk average \citep{Guzman2017}}
 \tablenotetext{d}{\citet{Bockelee2008}}
 \tablenotetext{e}{\citet{Arpigny2003}} 
\end{table*}

\subsubsection{Spin Temperature}

Hydrogen atoms have nuclear spin angular momentum, and molecules that contain two or more H atoms in symmetrical locations have quantum states distinguished by the total spin angular momentum ($I$). In the case of H$_2$O, for example, $I=1$ for ortho ($o$-H$_2$O) and 
$I=0$ for para ($p$-H$_2$O). Spontaneous conversion of spin state (e.g. ortho to para) is strongly forbidden by the quantum mechanical selection rules. The rotational energy levels for each spin state form a discrete ladder, and the energy of the lowest-lying level is lower for $p$-H$_2$O than $o$-H$_2$O by 34.2 K. The relative population of the ortho-to-para ratio (OPR) of H$_2$O is described as
\begin{equation}
{\rm OPR}= \frac{3\Sigma (2J+1)  \exp\left(\frac{-E_o(J_{K_a,K_c})}{k_{\rm B}T_{\rm spin}}\right)}{\Sigma (2J+1)  \exp\left(\frac{-E_p(J_{K_a,K_c})}{k_{\rm B}T_{\rm spin}}\right)},
\end{equation}
where the statistical weight of ortho and para state are 3 and 1, respectively, $T_{\rm spin}$ is spin temperature, and $K_a$ and $K_c$ are the quantum numbers to describe the projected angular momentum.
While the OPR is 3 in the high $T_{\rm spin}$ limit, the OPR is smaller than 3 due to the energy difference (34.2 K) of the lowest-lying state at $< 50$ K \citep{Hama2013}.

The OPR of H$_2$O, NH$_3$, and some other hydrated molecules have been measured in comets, assuming that $T_{\rm spin}$ is equal to the temperature of the formation site of the molecule \citep[e.g.][]{Mumma1987, Mumma2011,Shinnaka2020}. The derived spin temperature range is $20-60$ K with the typical value of $\sim 30$ K. For H$_2$O, the corresponding OPR is $\sim 2.5$ \citep{Mumma2011}.

In protoplanetary disks, on the other hand, the OPR of warm water vapor ($> 300$ K) detected in mid-infrared wavelength is consistent with OPR$=3$ \citep{Pontoppidan2010,vanDishoeck2013}. {\it Herschel} detected emission lines of cold water vapor ($<100$ K) in TW Hya; the OPR is $0.77 \pm 0.07$, which corresponds to $T_{\rm spin}= 13.5 \pm 0.5$ K.

The interpretation of $T_{\rm spin}$ as the temperature of the molecular formation site, however, turned out to be too simplistic. A laboratory experiment by \cite{Hama2018} showed that the OPR of H$_2$O desorbed from ice is 3 regardless of the temperature of the ice formation \citep[see also][]{Hama2016}. While the energy difference of the lowest-lying state between ortho and para is 34.2 K in the gas phase, the rotation motion is constrained and the energy difference becomes significantly small in the ice phase. Then the magnetic dipole interaction of a proton with neighboring protons enables rapid spin conversion \citep{Limbach2006,Hama2018}.

The lower OPR than the statistical value observed in comets and outer protoplanetary disks would thus be due to the gas-phase reactions; the spin state changes via proton exchange reactions such as $o$-H$_2$O + H$^+$ $\rightarrow$ $p$-H$_2$O + H$^+$. The cold water vapor in protoplanetary disks is much less abundant compared with the value expected from the desorption of ice (\S 3.1). The gas-phase formation of H$_2$O could thus play a role, while it is much less efficient than the grain-surface reactions at $<20$ K. 
As for comets, the effect of sublimation processes and gas-phase reactions in the coma needs to be considered for the interpretation of the observed OPR ratios \citep{Hama2013,Hama2016}

\section{Other Outstanding issues}

\subsection{From molecular clouds to disks}

It has long been discussed whether the molecular composition of volatiles in comets inherits ices in molecular clouds or is reset in the Solar nebula \citep[e.g.][]{Mumma2011}. To pursue this question, we need to investigate not only Class II disks but also the disk formation stage (i.e. Class 0-I) (section 1). Studies of forming disks have seen significant progress in the last decade.

On the theoretical side, disk formation has been studied by hydrodynamics simulations starting from the collapse of the parental cloud core, considering the effect of radiation transfer and magnetic fields to derive the physical structures (e.g. size, mass, temperatures) of the disks as a function of time \citep[e.g.][]{Machida2011,Tsukamoto2015, Hennebelle2016,tsukamoto23}.
The molecular evolution of gas and ice is investigated by calculating the chemical reaction network along the gas flow from the cloud core to the disk, considering the temporal variation of density, temperature, and UV radiation flux along the trajectories \citep[e.g.][]{Visser2009, Visser2011, Yoneda2016, drozdovskaya2016, Furuya2017, Aikawa2020}. Stable abundant ices such as H$_2$O and CO$_2$ are delivered to the disk without significant alteration, unless the trajectory is close to the outflow cavity where photodissociation is effective. [Outflows create a cavity (a pair of cone-shaped low-density regions) in the envelope. The wall of the outflow cavity could be irradiated by the UV radiation from the central object.] 
The abundance of COMs, on the other hand, could increase in the ice mantle as the thermal diffusion and reactions are activated in warm circumstellar conditions \citep[e.g.][]{garrod06,garrod22}.

Observations of forming disks are more complicated than those of Class II, since the emission from the disk needs to be distinguished from the infalling envelope and outflow. Careful analysis of spatial distributions and velocity structures shows that emission lines of rare isotopes of CO (with optical depth lower than their normal isotope), H$_2$CO, and CS tend to probe the disk, while SO traces week accretion shock onto the disk \citep{Sakai2014,Aso2017,Oya2018, Harsono2021,Tychoniec2021, Garufi2021}. The vertical stratification of molecules similar to that in Class II disks (\S 3.1) is also found \citep{Podio2020}. \citet{vantHoff2020} show that the young disks tend to be warmer than Class II disks, with the CO snowline at $\ge$ several tens of au, possibly indicating a higher accretion rate (\S 2.2). Disks with warped structures are also found \citep[e.g.][]{sakai2019,sai20}. Recent observations found large-scale (thousands of au) non-axisymmetric steamers onto young disks \citep{Yen2014,Pineda2020,Garufi2022}. Such a flow provides chemically fresh interstellar material (e.g. carbon chains) and angular momentum that could perturb the disk structure to enhance mixing.

\subsection{Estimation of volatile composition in solids}

At the time of writing this review, the direct observations of ices in Class II disks are relatively limited (\S 3.3), compared with the line observations of gaseous species. Here we overview and discuss how we can estimate the composition and evolution of volatiles in the solid phase based on gaseous observations.

Observations of fresh sublimates are the most straightforward to reveal the ice composition, and are possible during luminosity outbursts. The outbursts, which are caused by a sudden increase of mass accretion from the disk to the central star, are observed in Class I or the transition phase of Class I to Class II (\S \ref{sec:gas}). While the water snowline is typically located at $\sim $ a few au in Class II disks, it extends further away upon outburst. Around V883 Ori the dust continuum observation suggests that the water snowline is located at $\sim 40$ au \citep{Cieza2016}. V883 Ori is a FU Ori type star, for which the duration of the outburst phase is $\sim 100$ years. Since it is shorter than typical timescales of the gas-phase chemistry, we can observe fresh sublimates.
Various COMs (e.g. CH$_3$CHO and CH$_3$OCHO) are detected; their abundances relative to CH$_3$OH are comparable to those in comets \citep{Lee2019, Yamato24a}. 
HDO is recently detected as well (Tobin et al., 2023) (section 3.4.1).When the burst ceases, the COMs and water will again freeze on grain surfaces.

As described in \S 3.1, observations of Class II disks outside the CO snowline indicate that carbon and oxygen are depleted in the warm molecular layer, most probably due to the sedimentation and radial drift of ice-coated dust grains.
Observations of molecules such as C$_2$H suggest a high C/O ratio ($>1$) in the gas phase, which in turn suggests that the solid phase is oxygen-rich, e.g. with abundant water ice.
The molecular composition of ice could then be determined by the competition of the timescales of dust growth, settling, and molecular evolution in the upper layers of the disk. If the grain growth and settling are fast enough, the ices inherited from molecular clouds might be stored and survive in the midplane, which is cold and shielded from UV and X-ray radiation.
Interestingly, recent observations suggest that the growth and sedimentation of dust grains and CO depletion proceed over the short timescale from Class 0/I to II \citep{Bergner2020,Zhang2020,Harsono2021,Ohashi23}.

It should be noted, however,  that the freeze-out of gaseous molecules onto grain surfaces continues even after the dust sedimentation.
While the turbulence in disks is weaker than previously expected (\S 2.1), weak turbulence can bring molecules processed in the upper layers to the midplane, where they are frozen onto grains due to low temperature and high density (\S 3.1). 
Chemical reactions inside the ice mantle also continue after the sedimentation.
For example, impinging cosmic-ray particles cause radiolysis \citep{Shingledecker2018} and/or induce photochemistry \citep{Gredel1989}, although the flux of such energy source at the midplane is uncertain \citep{Cleeves2014, Aikawa21_MAPS}. As the grains migrate to inner radii, the gradual temperature rise enables the thermal diffusion and reactions of radicals inside cracks and pours of the ice mantle. \citet{Krijt2020} constructed numerical models that include pebble formation from small dust grains, radial drift of pebbles, diffusion of gas and dust, and simple gas-grain chemistry. They found that the heavy CO depletion observed in several disks is reproduced only in the model with the chemical conversion of CO to less volatile molecules and pebble sedimentation and radial drift. More recently, \citet{Furuya22a} performed calculations with a larger chemical network to show that the depletion of nitrogen is less severe than that of carbon.

Inside the CO snowline, CO abundance varies among objects \citep{Zhang2019,zhang21}. In some disks, the CO abundance goes back to the canonical abundance, while it does not in other disks. Assuming that the gas/dust mass ratio is 100, there are two possibilities in the latter case: (i) CO is converted to less volatile species, or (ii) CO ice is locked outside the CO snowline due to the planetesimal formation and/or dust trap at the local pressure maximum. For (i), the product molecules would eventually sublimate when the ice-coated grains reach the inner warmer regions.
For (ii), on the other hand, the elemental abundances of carbon and oxygen would be low even in the inner hot regions, ultimately at the inner edge of the disk and in the accreting material from the disk to the central star.
\citet{Bosman2019} and \citet{McClure2020} observed gas at the innermost radius of the TW Hya disk to find that both carbon and oxygen are depleted. This method works for other elements as well. \citet{Kama2019} analyzed the stellar spectra of young disk-hosting stars to find depletion of Sulphur, which suggests that $89\pm 8 \%$ of elemental sulfur is in refractory form, e.g. FeS, and trapped in dust traps.
The radial variation of elemental abundances should depend on the relative positions of snowlines of major ices and the dust trap or planetesimal formation. Ideally, observations of gases inside the possible comet-forming dust ring could tell us the amount of volatiles used to make comets.

\section{Brief summary and future directions}\label{s:future}

Studies on Solar system material, including comets, and protoplanetary disks play complementary roles in exploring the formation of the Solar system. While comets and asteroids are remnants or fragments of planetesimals, i.e. the building blocks of planetary systems, we can observe their ingredients, i.e. molecules, dust grains, and ices in disks. While the chemical composition of the Solar system material contains rich but degenerated information about their formation environment, the statistical and spatially resolved observations of disks, including those around Class 0 and I objects, reveal the physical and chemical structures and their evolution. Variations of exoplanetary systems suggest that not all disks produce planetary systems like the Solar system. Understanding basic physical and chemical processes in disks is essential to reveal the specific conditions for the Solar system formation.

While ALMA has revealed the thermal emission of dust with sub-arcsecond resolution ($\sim 5$ au), derivation of the mass distribution of dust grains depends on the opacity, which, in turn, depends on the size distribution of grains. The combination of the observations at various wavelengths, polarization, and theoretical studies on dust properties are crucial (\S 2).
In order to probe the dust in the inner radius, which tends to be optically thick even in millimeter wavelength, low-frequency observations by Band 1 ALMA and ngVLA are essential. These low-frequency instruments are also suited for observations of major volatile molecules such as NH$_3$ and CH$_3$OH. In recent years, various ring and linear molecules are newly detected in the line surveys of molecular cloud TMC-1 at Green Bank Telescope and Yabe 40 m telescope at $\le 50$ GHz \citep{Cernicharo2020, McGuire2020}. While the detection of new molecules in disks could be limited by sensitivity, since large molecules suitable for low-frequency observations tend to be frozen in ice mantle, disks in outburst could be a good target.

Molecular line observations by ALMA are revealing the gas dynamics (\S 2), and composition and isotope ratios of disk gas (\S 3). Theoretical studies of disk chemistry and multi wavelengths observations are crucial to estimate the composition of solids, which is directly linked to comets (\S3 and \S4). It is clear that a significant fraction of dust grains are grown and decoupled from gas in Class II disks, which makes the elemental abundance heterogeneous. Combined analysis of various molecular lines is important to derive not only the chemical composition but also gas mass distributions in disks. One of the key probes of gas mass is the far-infrared line of HD (\S 3.1), which would be observable with a new instrument on a suborbital platform, such as a balloon,
or a future far-infrared space mission \citep{Bergin19}.

In shorter wavelengths, the James Webb Space Telescope (JWST) is now providing a comprehensive view of inner disk chemistry in protoplanetary disks around stars of a wide range of masses. 
Active programs include the JWST Disk Infrared Spectroscopic Chemistry Survey \citep[JDISCS,][]{Pontoppidan24}, and the Mid-infrared Disk Survey \citep[MINDS,][]{Kamp23}.
These surveys promise to yield inventories of a wide range of bulk volatile species and their isotopologues within the snowline, including water, organics, and nitrogen-bearing species. While the JWST observations will be able to survey disks around stars of all masses, its medium-resolution spectrometer will not be able to spectrally resolve individual lines. However, a new generation of high-resolution spectrometers on ground-based facilities will continue to offer complementary spectroscopy at high resolution. In particular the new CRIRES+ instrument \citep{Dorn14} on the European Southern Observatory Very Large Telescope (ESO-VLT), and in the future, the METIS instrument \citep{Brandl10} of the European Extremely Large Telescope and MICHI \citep{Packham18} on the Thirty Meter Telescope will provide sensitive 3-12\,$\mu$m spectroscopy of protoplanetary disk chemistry at high spectral and spatial resolution. 
Finally, the US 2020 Decadal Survey recommended either a far-infrared or X-ray probe class misssion in the 2030s, and several concepts to trace the physics and chemistry of volatiles around the snowline using far-infrared spectroscopy are under consideration \citep[e.g., PRIMA,][]{Glenn23}.


%

\vskip .5in
\noindent \textbf{Acknowledgments.} \\
Y.A. acknowledges support by NAOJ ALMA Scientific Research Grant Numbers 2019-13B, Grant-in-Aid for Scientific Research (S) 18H05222, (B) 24K00674, and Grant-in-Aid for Transformative Research Areas (A) 20H05844 and 20H05847. 
S.~O. is supported by JSPS KAKENHI Grant Numbers JP18H05438 and JP20H00182. Part of this work was carried out at the Jet Propulsion Laboratory, California Institute of Technology, under a contract with the National Aeronautics and Space Administration (80NM0018D0004).

\bibliographystyle{sss-three.bst}
\bibliography{refs.bib}

\end{document}